\RequirePackage{fix-cm}
\documentclass{svjour3}       
\smartqed  
\usepackage{appendix}
\usepackage{amsmath,amsfonts,amssymb}
\usepackage{graphicx}
\usepackage{lineno}
\usepackage{array}
\usepackage{longtable}
\usepackage{natbib}
\setcitestyle{aysep={}}
\usepackage{adjustbox}
\usepackage[figuresright]{rotating}
\usepackage[hidelinks]{hyperref}
\newcommand*\patchAmsMathEnvironmentForLineno[1]{%
\expandafter\let\csname old#1\expandafter\endcsname\csname #1\endcsname
\expandafter\let\csname oldend#1\expandafter\endcsname\csname end#1\endcsname
\renewenvironment{#1}%
{\linenomath\csname old#1\endcsname}%
{\csname oldend#1\endcsname\endlinenomath}}%
\newcommand*\patchBothAmsMathEnvironmentsForLineno[1]{%
\patchAmsMathEnvironmentForLineno{#1}%
\patchAmsMathEnvironmentForLineno{#1*}}%
\AtBeginDocument{%
\patchBothAmsMathEnvironmentsForLineno{equation}%
\patchBothAmsMathEnvironmentsForLineno{align}%
\patchBothAmsMathEnvironmentsForLineno{flalign}%
\patchBothAmsMathEnvironmentsForLineno{alignat}%
\patchBothAmsMathEnvironmentsForLineno{gather}%
\patchBothAmsMathEnvironmentsForLineno{multline}%
}

\begin{document}

\title{Stable Boundary Layers with Subsidence: Scaling and Similarity of the Truly Steady State }

\author{Thijs Bon         \and
        Raúl Bayoán Cal \and Johan Meyers
}

\institute{T. Bon \at
              Department of Mechanical Engineering, KU Leuven, Celestijnenlaan 300A-Bus 2421, B3001 Leuven \\
              \email{thijs.bon@kuleuven.be}           
           \and
           R. B. Cal \at
           Portland State University, Department of Mechanical and Materials Engineering, Portland, Oregon 97207, USA \\
            \and
            J. Meyers \at
            Department of Mechanical Engineering, KU Leuven, Celestijnenlaan 300A-Bus 2421, B3001 Leuven
}



\date{Received: DD Month YEAR / Accepted: DD Month YEAR}

\maketitle

\begin{abstract}
The stable boundary layer (SBL) subjected to large-scale subsidence is studied through large-eddy simulations (LESs) with fixed surface temperature and a linear subsidence velocity profile. These boundary layers reach a truly steady state, where thermal equilibrium is established by a balance between surface cooling and subsidence-induced heating. We identify three governing dimensionless groups by scaling the governing equations with the geostrophic wind and Coriolis frequency, and systematically investigate the impact of these external parameters on global flow properties and mean profiles in the steady state.
The SBL depth, low-level jet, and the magnitude of the turbulent momentum flux are reduced when the subsidence rate or Buoyancy number increases, while surface heat flux is enhanced. The shape of normalized mean profiles of temperature and heat flux is mainly determined by the subsidence rate, while they collapse for varying buoyancy and surface Rossby numbers. We develop empirical correlations for the stability parameter $h_\theta/L_O$ and a thermal shape factor, and propose a new unidirectional geostrophic drag law, to form a closed set of equations that estimates relevant flow properties from external parameters. The estimation errors compared to the LES data are less than 5\% for friction velocity and surface heat flux, and at most 10\% for the SBL depth $h_\theta$. Within the surface layer, dimensionless velocity and temperature gradients in the steady SBL with subsidence show acceptable agreement to Monin--Obukhov similarity theory, while the collapse is improved when a recently proposed mixed scaling parameter, that includes $h_\theta/L_O$, is used. 
\end{abstract}

\section{Introduction}
\label{sec:intro}
One of the key challenges in atmospheric and environmental sciences is understanding of the stable boundary layer (SBL), which is relevant to the renewable energy sector, air quality and pollutant dispersion, numerical weather prediction and climate modelling {(e.g. \cite{Steeneveld2014}, \cite{McWilliams2023}, and references therein)}. The atmospheric boundary layer (ABL) commonly exhibits stable stratification during nighttime over land (also referred to as the nocturnal boundary layer or NBL), due to radiative cooling of the surface. SBLs can further be generated by advection of warm air over a colder surface, for example in maritime areas or in the polar regions. The main processes that govern the structure of the SBL are radiation, turbulence, horizontal advection and subsidence \citep{Stull1988, Wyngaard2010}. 

Subsidence, large-scale downward movement of air, and as a consequence near-surface flow divergence, is often associated with synoptic high-pressure systems, fair weather conditions and clear skies \citep{Carlson1985}. Furthermore, polar areas experience persistent subsidence due to planetary-scale circulations \citep{Holton2004}. Motivated by the relatively small magnitude of these vertical motions compared to horizontal velocity components and the difficulty to accurately measure them, subsidence is often neglected in numerical investigations of the SBL \citep{Mirocha2010, Stoll2020}. 
{However}, early observational studies already concluded that the heating rate associated with subsidence can be as large as the cooling rates due to turbulence and radiation \citep{Carlson1985, Mirocha2005a}, and should therefore not be neglected. In the present study, we investigate the SBL under influence of subsidence and the related steady state regime through an extensive set of large-eddy simulations (LESs).
Although effects of subsidence on the SBL have been investigated previously, a systematic analysis of the full parameter space that plays a role in the SBL with subsidence has not been performed before.

In a stably stratified environment, subsidence brings down warm air from aloft, thereby reducing the boundary-layer height and affecting the atmospheric heat balance \citep{Stull1988}. This can affect many other physical processes, such as dispersion of pollutants \citep{Shi2022}, and growth/ablation of a snow/ice pack due to changes in surface heat flux \citep{Mirocha2010}. Moreover, the formation of  marine boundary layer clouds, which poses a significant source of uncertainty in climate models, is influenced by subsidence \citep{Bellon2012, Chung2012}. Prior work has demonstrated that incorporating atmospheric subsidence into numerical simulations improves the agreement with measurement data of the SBL over the Arctic Ocean \citep{Mirocha2010} and the Antarctic Plateau \citep{Vignon2017a, VanderLinden2019}. Other studies highlight the importance of including subsidence in modelling of the unstable boundary layer over both land \citep{Blay-Carreras2014a, Rey-Sanchez2021} and sea \citep{Mazzitelli2014}. 

The LES study of \cite{Mirocha2010}, which was based on observations from the Surface Heat Budget of the Arctic Ocean field experiment (SHEBA), showed that the addition of subsidence reduces vertical momentum fluxes while enhancing vertical heat flux, increases temperature gradients, and limits the growth of the boundary layer, leading to a ``nearly  steady boundary layer behaviour during periods of constant forcing''. More recently, L19 performed centimetre-scale LESs of a weakly and very stable boundary layer and compared results to observational data from the Dome C station on the Antarctic Plateau, where the absence of the diurnal cycle during the Antarctic winter facilitated the examination of a long-lived SBL. They concluded that the inclusion of subsidence can result in a truly steady SBL, where the turbulent cooling towards the surface is balanced by the heating due to subsidence. This statistically steady state, characterized by a thermal equilibrium (see section \ref{subsec:thermal_equilibrium}), is the main focus in the present study. We note that the term ``equilibrium'' here should not be confused with equilibrium in developing turbulent boundary layers, which describes a situation where the normalized flow properties do not longer depend on streamwise position \citep{Devenport2022}.

In typical investigations of the SBL through LES or direct numerical simulation (DNS), a true steady state is never reached. Many of these studies are based on the well-documented case setup of the Global Energy and Water Cycle Experiment Atmospheric Boundary Layer Study 1 (GABLS1, \cite{Beare2006}), which currently is still used as a reference case to e.g., validate new LES codes \citep{VanHeerwaarden2017}, evaluate new subgrid models or surface parametrizations \citep{Matheou2014,Maronga2020, Dai2021} and assess grid-size sensitivity of LES models \citep{Maronga2022}. Slightly modified configurations have been used to study effects of increasing stability \citep{Huang2013, Sullivan2016a, Narasimhan2023}, heterogeneous surface temperatures \citep{Stoll2009, Mironov2016}, or interactions between the SBL and wind farms \citep{Allaerts2018, Strickland2022a}. The GABLS1 setup is characterized by a constant cooling rate as the lower boundary condition, meaning that the surface temperature keeps decreasing in time. In studies that employ DNS to study the SBL, there is more variety in the surface temperature boundary condition, involving either a prescribed surface heat flux \citep[e.g.][]{Nieuwstadt2005, Flores2011, Gohari2017}, fixed bottom temperature \citep{Ansorge2014, Shah2014}, or a combination thereof \citep{Gohari2018}. Because the surface cooling is not compensated by any heat source, none of the simulations in the aforementioned studies reaches a truly steady state, where both the turbulent and thermal energy of the system are in (statistical) equilibrium. The ``quasi-steady'' state that the SBL reaches after a finite time is commonly employed, in which the mean profiles of velocity and turbulence statistics remain constant while only the temperature profile keeps changing. This quasi-steady period is then used to analyze results and compare cases, while the domain-averaged temperature still declines (see Eq. \eqref{eq:balance}). In numerous applications however, it is desired to have long averaging times under truly stationary conditions, without any transient effects, to permit more systematic analysis and comparison of multiple different simulations. For example, the study of large-scale secondary motions that can be induced by heterogeneous surface roughness or temperatures, requires long averaging times \citep{Bon2022, Castro2021, Schafer2022a}. When employing LES for a systematic analysis of the effects of large wind farms on atmospheric heat fluxes, examination of stationary statistics considerably simplifies understanding \citep{Sescu2014, Maas2022}. Moreover, in assessments of the sensitivity of an LES code to the grid spacing or the subgrid-scale model, constantly decreasing surface temperature in typical SBL simulations prevents using long averaging times \citep{Maronga2022}.

In the next section (\ref{subsec:thermal_equilibrium}) , we start with a brief description of thermal equilibrium in the SBL with subsidence. The LES code and numerical methodology are subsequently outlined in Sect. \ref{subsec:numericalmethods}. Based on dimensional analysis, we identify the relevant non-dimensional groups in Sect. \ref{subsec:dimensionalanalysis}, and perform a series of LESs where the surface temperature, subsidence rate and surface roughness are varied. The case setup can be roughly viewed as a combination of the GABLS1 case and the configuration used by L19, as further detailed in Sect. \ref{subsec:Simulations}. Based on the LES results, we examine how the steady state depends on the relevant external parameters in Sect. \ref{sec:results}. Different methods of estimating the SBL depth are compared in Sect. \ref{subsec:sbl_depth}, followed by an investigation of scaled mean profiles (Sect. \ref{subsec:mean_profile_scaling}). In Sect. \ref{subsec:GDL}, a geostrophic drag law is proposed for the steady SBL with subsidence, and present model equations that accurately estimate global flow properties of interest based on external parameters. An assessment of the validity of the widely used Monin--Obukhov similarity theory (MOST) in the presence of subsidence is given in Sect. \ref{subsec:MOST}. Finally, the main conclusions are summarized in Sect. \ref{sec:conclusions}.

\section{Theory and Methods}
\label{sec:theory_methods}

\subsection{Conditions for Thermal Equilibrium}
\label{subsec:thermal_equilibrium}
In typical idealized LES, subsidence and radiation effects are not taken into account, such that the temporal evolution of the potential temperature $\theta$ is determined by the divergence of the turbulent heat flux $\partial q/\partial x_i$ (see also Eq. \eqref{eq:temperature_equation} further below). Taking the horizontal average and integrating vertically, we find:
\begin{equation}
	\int_{0}^{h}\frac{\partial \langle  \theta \rangle}{\partial t} \mathrm{d}z = -\int_{0}^{h}\frac{\partial \langle q_z \rangle}{\partial z} \mathrm{d}z = \langle q_{z}  \rangle_0 \equiv q_0 < 0
	\label{eq:balance}
\end{equation}
where $z$ denotes the vertical coordinate, angular brackets denote horizontal averaging, $\langle q_z \rangle = \langle {w'\theta'} \rangle + \langle q_z^{\mathit{sgs}}  \rangle$ is the total (turbulent + subgrid-scale) heat flux, $h$ can be any height above which $q_z$ is assumed to vanish (i.e. $\langle q_z \rangle_h =0$), and $q_0$ is the surface heat flux. Note that the order of the integral and the derivative on the left-hand side can be changed provided that $h$ remains constant. Since the surface heat flux in an SBL is per definition non-zero and negative, Eq. \eqref{eq:balance} shows that the mean temperature in the boundary layer will keep decreasing if there are no additional heat sources.

\begin{figure}[h]
	\centering\includegraphics[trim=0cm 0.2cm 0cm 0cm, clip]{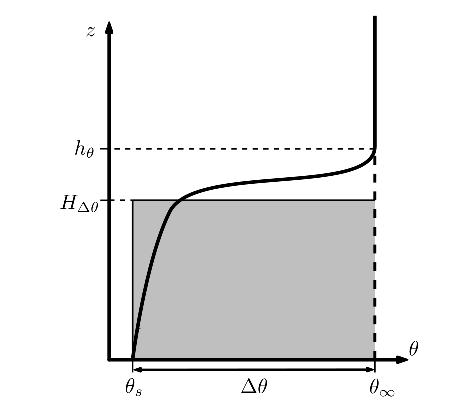}
	\caption{Schematic diagram of SBL temperature profile with geometrical definitions of thermal boundary-layer depth $h_\theta$, integral depth scale $H_{\mathrm{\Delta}\theta}$ and global temperature difference $\mathrm{\Delta}\theta$. The shaded area within the rectangle $H_{\mathrm{\Delta}\theta}\mathrm{\Delta}\theta$ equal the area under the potential temperature curve. Figure inspired by Fig. 12.4 in \cite{Stull1988}  }
	\label{fig:schematic}
\end{figure}

This changes when a subsidence term of the form $Q_s = -w_s(z) (\partial \langle\theta\rangle /\partial z$) is added to the temperature equation, acting as a heat source. A number of different parametrizations for the subsidence velocity profile $w_s(z)$ have been proposed in previous studies, such as piecewise linear \citep{VanderLinden2019, Maas2022}, cubic \citep{Mirocha2010}, polynomial \citep{Mazzitelli2014} or exponential \citep{Bellon2012} shapes.
In order to minimize the number of external parameters, we adopt a simple linear subsidence profile $w_s(z) = -\mathrm{\Gamma}_w z$ (with $\mathrm{\Gamma}_w > 0$), similar to e.g. \cite{Chung2012}. Note that due to continuity, the subsidence rate $\mathrm{\Gamma}_w$ directly corresponds to the magnitude of the associated large-scale horizontal divergence, which is thus assumed constant with height \citep{Mazzitelli2014}.  

Including the subsidence term, the horizontally-averaged temperature equation for the steady state can be written as:
\begin{equation}
	\frac{\partial \langle \theta \rangle }{\partial t} = 0 = -\frac{\partial \langle q_z \rangle }{\partial z} + \mathrm{\Gamma}_w z \frac{\partial \langle \theta \rangle }{\partial z},
	\label{eq:balance2}
\end{equation}
showing that the flow is in thermal equilibrium if cooling by vertical heat-flux divergence is equal to heating by subsidence (L19). An important implication of Eq. \eqref{eq:balance2} is that thermal equilibrium can only be achieved if the non-turbulent free atmosphere or residual layer is unstratified, because the turbulent heat flux $q_z=0$ there. 

Performing the same integration as in Eq. \eqref{eq:balance} then leads to:
\begin{equation}
	q_0 = -\mathrm{\Gamma}_w \int_{0}^{h}z \frac{\partial \langle \theta \rangle }{\partial z} \mathrm{d}z = -\mathrm{\Gamma}_w \left( z \langle\theta\rangle |_0^h - \int_{0}^{h} \langle \theta \rangle \mathrm{d}z  \right)
	\label{eq:partial_integration}
\end{equation}
where the last equality follows from partial integration. As usual in LES of the SBL, we only consider dry air under the Boussinesq-approximation, allowing us to change the reference temperature by a constant. Furthermore, following Fig. \ref{fig:schematic}, it is assumed that $h=h_\theta$ is the thermal boundary-layer height above, where the temperature remains equal to the constant free-atmosphere temperature ($\theta(z\geq h_\theta)=\theta_\infty$) and the surface temperature $\theta_s = \theta_\infty - \mathrm{\Delta}\theta$ is fixed, such that the substitution $\theta$ by a temperature deviation $\theta_\delta = \theta - \theta_\infty$ is made without loss of generality. Consequently, the first term on the right-hand side of Eq. \eqref{eq:partial_integration} vanishes and obtain:
\begin{equation}
	q_0 = \mathrm{\Gamma}_w \int_{0}^{h_\theta} \langle \theta_\delta \rangle \mathrm{d}z = \mathrm{\Gamma}_w\mathrm{\Delta}\theta H_{\mathrm{\Delta}\theta}.
	\label{eq:q0}
\end{equation}
The integral of the temperature deviation can be interpreted as the `accumulated cooling', and is alternatively written as $\mathrm{\Delta}\theta H_{\mathrm{\Delta}\theta}$, where $H_{\mathrm{\Delta}\theta} = 1/\mathrm{\Delta}\theta \int_{0}^{h_\theta} \langle \theta_\delta \rangle \mathrm{d}z $ is the integral depth scale of the SBL (see Fig. \ref{fig:schematic} and p. 504 in \cite{Stull1988}). It is noted that, in the community that studies developing thermal boundary layers, $H_{\mathrm{\Delta}\theta}$ is referred to as the thermal displacement thickness \citep{Wang2003}. 

Equation \eqref{eq:q0} shows that the SBL in our simplified framework reaches a steady state when the surface heat flux (left-hand side) is equal to the bulk heat flux due to subsidence heating (right-hand side), calculated as the integrated temperature deviation multiplied by the linear subsidence rate \citep[p. 515]{Stull1988}. Moreover, Eq. \eqref{eq:balance2} shows that there is an explicit relation between mean profiles of temperature and turbulent heat flux. However, as pointed out by L19, the internal distribution of heat flux and temperature in the boundary layer, and likewise the surface heat flux, (thermal) boundary-layer height and mean temperature, remain unknown a priori. The set of simulations that we perform aims to shed light on relations between these parameters and assess similarity of mean profiles, as discussed in Sect. \ref{subsec:mean_profile_scaling}.

Heuristically, one can assume that the integral $\int_{0}^{h_\theta} \langle \theta_\delta \rangle \mathrm{d}z$ is proportional to $h_\theta\mathrm{\Delta}\theta$, with a proportionality parameter $S_\theta$ that depends on the shape of the temperature profile. This ``shape factor'' is then defined as:
\begin{equation}
	S_\theta = \frac{1}{h_\theta\mathrm{\Delta}\theta}\int_{0}^{h_\theta} \langle \theta_\delta \rangle \mathrm{d}z = \frac{H_{\mathrm{\Delta}\theta}}{h_\theta}  =  \frac{q_0}{\mathrm{\Gamma}_w h_\theta \mathrm{\Delta}\theta}
	\label{eq:shape_factor}
\end{equation}
where the last equality only holds in thermal equilibrium and follows from Eq. \eqref{eq:q0}. Note that this same parameter also appears in Eq. \eqref{eq:balance2} if $q_z$, $\theta$ and $z$ are scaled with $q_0, \mathrm{\Delta}\theta$ and $h_\theta$. The denominator $(\mathrm{\Gamma}_w h_\theta \mathrm{\Delta}\theta)$ could be interpreted as the maximal possible bulk heat flux due to subsidence, that would occur if all subsidence heating would take place at an infinitely thin temperature inversion layer of strength $\mathrm{\Delta}\theta$ at height $h_\theta$. Hence, $S_\theta$ can also be explained as the ratio of the actual bulk subsidence heat flux (see Eq. \eqref{eq:q0}) and the theoretical maximum, or the `efficiency' of the subsidence heating. Evidently, $S_\theta = 1/2$ in case of a linear temperature profile between $z=0$ and $z=h_\theta$, while in general $S_\theta$ should be between 0 and 1 (see also Fig. \ref{fig:schematic} here and Fig. 12.5 in \cite{Stull1988}). 
In Sect. \ref{subsec:mean_profile_scaling} we will investigate how $S_\theta$ is affected by external simulation parameters. 
\subsection{Numerical Methodology}
\label{subsec:numericalmethods}
The governing equations of our LES model are filtered versions of the continuity, Navier-Stokes and potential-temperature equations for incompressible flow under the Boussinesq approximation:
\begin{equation}
	\frac{\partial {u}_i}{\partial x_i} = 0, \label{eq:continuity}
\end{equation}
\begin{equation}
		\frac{\partial {u}_i}{\partial t}  =  -\frac{\partial}{\partial x_j} ({u}_i {u}_j + \tau_{ij}^{\mathit{sgs}} ) - \frac{\partial {p}^\star}{\partial x_i} \\ + f_c\epsilon_{ij3}({u}_j - G \delta_{j1}) +   \frac{g}{\theta_r} ({\theta}_\delta - \langle {\theta}_\delta \rangle) \delta_{i3}, 	\label{eq:momentum_equation} 
\end{equation}
\begin{equation}
		\frac{\partial {\theta}_\delta}{\partial t} = - \frac{\partial}{\partial x_j} ( {u}_j  {\theta}_\delta + q_j^{\mathit{sgs}}) + \mathrm{\Gamma}_w x_3 \frac{\partial \langle {\theta}_\delta \rangle}{\partial x_3},
	\label{eq:temperature_equation}
\end{equation}
where indices $i=1,2,3$ correspond to respective horizontal ($x,y$) and vertical ($z$) directions, ${u}_i$ are the filtered velocity components and ${\theta}_\delta = {\theta}-\theta_\infty$ is the filtered temperature deviation from the top temperature. The boundary layer is subjected to a large-scale pressure forcing, determined by the Coriolis frequency $f_c = 1.39\times 10^{-4}$ s$^{-1}$ and the geostrophic wind speed $G=8$ m s$^{-1}$. The gravitational acceleration $g=9.81$ m/s$^2$, and the reference temperature $\theta_r=\theta_\infty=263.5$ K. These values are equal across all simulations and align with those in the GABLS1 case \citep{Beare2006}). As described in the previous section, $\mathrm{\Gamma}_w$ is the subsidence rate, i.e. the gradient of the subsidence velocity profile in $\mathrm{s}^{-1}$. Following L19, we use the horizontally averaged vertical temperature gradient rather than the local one, and neglect horizontal velocity divergence in the momentum equation (see Appendix 1 in their work for justification). 
The subgrid-scale (SGS) fluxes $\tau_{ij}^{sgs}$ and $q_j^{sgs}$ are modelled with the turbulent kinetic energy (TKE) model, based on Deardorff (1980), which may be the most widely used SGS model in LES \citep{Gibbs2016, Stoll2020}. The modified filtered pressure ${p}^\star$ is in fact the deviation from the background pressure, divided by a reference density, and includes the trace of the subgrid stress tensor \citep{Allaerts2015}. 

The governing equations \eqref{eq:continuity}-\eqref{eq:temperature_equation} are solved using the SP-Wind solver, an in-house DNS/LES code developed over the past 15 years at KU Leuven. It has been widely used to study the interaction between wind-farms and the (stably) stratified Ekman layer \citep{Goit2015, Allaerts2016, Allaerts2018, Lanzilao2024}, as well as stably stratified flows over thermally inhomogeneous surfaces \citep{Bon2023}. In SP-Wind, time integration is performed using a classical fourth-order Runge--Kutta scheme with a fixed Courant--Friedrichs--Lewy (CFL) number of 0.4. The code employs pseudo-spectral discretization in the horizontal directions with the 3/2-dealiasing rule for the nonlinear terms, and a fourth-order energy conserving finite-difference scheme for the vertical direction \citep{Verstappen2003}. Consequently, the horizontal boundary conditions are periodic, while at the top of the domain symmetry conditions for horizontal velocity components and potential temperature ($(\partial({u},{v},{\theta})/\partial z =0$) are applied, and the vertical velocity (${w}$) is 0.

At the lower boundary of the domain, we apply MOST to model the surface fluxes for momentum and temperature. A note on using MOST as a boundary condition is presented in Appendix 3. 
The surface flux parametrization in MOST depends on the roughness lengths for momentum ($z_{0m}$) and heat ($z_{0h}$), and  the fixed surface temperature $\theta_s=\theta_\infty -\mathrm{\Delta}\theta$. In order to limit our parameter space, we take $z_{0h} = z_{0m} = z_0$. 
For further details on the SP-wind code, implementation of MOST and the TKE subgrid model, we refer to \cite{Allaerts2016} and \cite{Allaerts2018}. The latest version of SP-Wind employs a full 3-dimensional domain decomposition for efficient parallel computation, allowing to run the majority of simulations on 12 nodes, each containing 128 AMD Epyc 7763 cores in the HPC cluster of the Flemish Supercomputing Center (VSC).

\subsection{Dimensional Analysis}
\label{subsec:dimensionalanalysis}
In the idealized LES of stably stratified boundary layers with subsidence  considered, there are at least five relevant external parameters. Provided that the top of the computational domain is high above the turbulent boundary layer and the surface temperature is fixed, both the top temperature $\theta_\infty$ and the surface temperature $\theta_s$ remain constant. Therefore, the global buoyancy difference $ (g/\theta_r)(\theta_t-\theta_s)= (g/\theta_r)\mathrm{\Delta}\theta \equiv \mathrm{\Delta} B $ is one input to our simulations. With that, the full set of governing parameters is $\{G, f_c, z_0, \mathrm{\Delta} B, \mathrm{\Gamma}_w\}$. Applying the Buckingham-Pi theorem to these five parameters which involve only two dimensions (length and time), three non-dimensional $\mathrm{\Pi}$-groups are formed:

\begin{equation}
	\mathit{Ro_0} = \frac{G}{f_c z_0}, \quad \mathit{Bu} = \frac{\mathrm{\Delta} B}{G f_c},\;\; \mathrm{and}\;\; \mathrm{\Pi}_w = \frac{\mathrm{\Gamma}_w}{f_c}.
	\label{eq:parameters}
\end{equation}

The first group is a Rossby number based on the surface roughness length $z_0$ and geostrophic wind $G$, commonly referred to as the surface Rossby number \citep[e.g.][]{Tennekes1973, Zilitinkevich2002a}. The second is the Buoyancy number ${Bu}$, a global stability parameter that characterizes thermal stratification across an Ekman layer \citep{Csanady1974, Swinbank1974}. It is a ratio between buoyancy forces and Coriolis forces, or can be viewed alternatively as a product of a Richardson number $\mathit{Ri} = \mathrm{\Delta} B h/G^2$ and a Rossby number $\mathit{Ro_h} = G/(fh)$, where $h$ would be the boundary-layer height. The last group, $\mathrm{\Pi}_w$, determines the magnitude of the subsidence and can be interpreted as the ratio between the timescales of subsidence heating and Coriolis force. Note that this is a purely practical choice of scaling the subsidence with one of the external parameters, since $f_c$ and $\mathrm{\Gamma}_w$ do not appear in the same equation (see Eqs. \eqref{eq:continuity}-\eqref{eq:temperature_equation}) 

\subsection{Suite of Simulations}
\label{subsec:Simulations}
Based on the dimensionless groups in Eq. \eqref{eq:parameters}, three sets of simulations are performed where one of the parameters is changed while the others are kept at a reference value. As mentioned above, $G$, $f_c$ and $g/\theta_r$ are held constant over all simulations, thus only $z_0$, $\mathrm{\Delta} \theta$ and $\mathrm{\Gamma}_w$ are varied. These three simulation sets are referred with letters `Z', `T' and `S', respectively. The simulation with the `default' values $z_0 = 10^{-3}$ m ($\mathit{Ro_0}=5.8 \times 10^{7}$), $\mathrm{\Delta}\theta = 3$ K ($\mathit{Bu} = 100$) and $\mathrm{\Gamma}_w = 1.25 \times 10^{-5}$ s$^{-1}$ ($\mathrm{\Pi}_w = 0.09$) is referred to as `S3T2Z2'. An overview of all simulations is provided in Table \ref{tab:overview}. 

In set `Z', the surface roughness is varied between $10^{-4}$ m, as typically used to represent sea surface \citep{Liu2021a, Lanzilao2024} and $10^{-1}$ m which was used in the GABLS1 case to represent onshore conditions \citep{Beare2006}. The reference value of $10^{-3}$ m lies in between, and corresponds to that used by L19 to represent the Dome C site in Antarctica. In recent LES studies of SBLs \citep{VanderLinden2019, Couvreux2020, Narasimhan2023}, the roughness length for heat is one order of magnitude smaller than for momentum, as also recommended by \cite{Vignon2017} based on measurements in polar regions. However, \cite{Couvreux2020} report little sensitivity of simulation results to changing $z_{0h}$ from $10^{-4}$ m to $10^{-3}$ m, and confirmed through preliminary simulations.

The range of temperature differences in set `T' spans from $\mathrm{\Delta}\theta=1.5$ K to 12 K. The `base' temperature difference of 3 K roughly corresponds to the difference between the temperature at the surface and the top of the inversion layer in the GABLS1 case, allowing direct comparison of mean profiles to the present simulations (see Fig. \ref{fig:profiles}k-o)

In set `S', the subsidence rate $\mathrm{\Gamma}_w$ is varied over two orders of magnitude, from $2.5 \times 10^{-6}$ s$^{-1}$ to $2\times10^{-4}$ s$^{-1}$, while the cases in the other sets have the base value of $1.25\times10^{-5}$ s$^{-1}$. To put these values into context, \cite{Carlson1985} estimated horizontal divergence rates of $0.09-3.3 \times 10^{-5}$ s$^{-1}$ based on a field experiment in Oklahoma. The majority of aforementioned LES studies that include subsidence, assumed a subsidence rate or horizontal divergence in the order of $10^{-6} - 10^{-5}$ s$^{-1}$. In particular, L19 essentially employs a linear subsidence velocity profile with gradient $4 \times 10^{-5}$ s$^{-1}$ up to 100 m above the surface, inferred from ERA-Interim model reanalysis and simulations from \cite{Baas2019}.

\begin{sidewaystable}
	\centering
	\caption{Overview of simulations and bulk quantities. The external parameters are subsidence rate $\mathrm{\Gamma}_w$, global temperature difference $\mathrm{\Delta}\theta$, surface roughness $z_0$, and the corresponding dimensionless subsidence, buoyancy and surface Rossby numbers ($\mathrm{\Pi}_w, \mathit{Bu}, \mathit{Ro_0}$ as defined in \eqref{eq:parameters}). Internal parameters are friction velocity $u_*$, kinematic surface heat flux $q_0$, thermal boundary-layer height $h_\theta$, Obukhov length $L_O$. The total runtime $t_r$  of each simulation is reported in the last column. Note that the last two rows are simulations by L19, derived from their Table 2}
	\label{tab:overview}
	\begin{tabular}{lllllll|llllll}
		\hline
		Case      & $\mathrm{\Gamma}_w$ (s$^{-1}$) & $\mathrm{\Delta}\theta$ (K) & $z_0$ (m)    & $\mathrm{\Pi}_w$ & $\mathit{Bu}$ & $\mathit{Ro}_0 $ & $u_*$ (m s$^{-1}$) & $q_0$ (K m s$^{-1}$)  & $h_\theta$ (m) & $L_O$ (m) &$t_{r}$ (h)& $R_q$ (\%) \\
		\hline
		S1        & 2.5 $\times 10^{-6}$                      & 3              & $10^{-3}$ & 0.02    & 100  & 5.8 $\times 10^7$      & 0.232& -1.31$\times10^{-3}$& 190& 644&84& 0.05         \\
		S2        & 6.25 $\times 10^{-6}$                     & 3              & $10^{-3}$ & 0.04    & 100  & 5.8 $\times 10^7$      & 0.223& -2.28$\times10^{-3}$& 152& 327&60& 0.66          \\
		S3T2Z2    & 1.25 $\times 10^{-5}$                     & 3              & $10^{-3}$ & 0.09    & 100  & 5.8 $\times 10^7$      & 0.212& -3.39$\times10^{-3}$& 134& 188&60&0.04          \\
		S4        & 2.5  $\times 10^{-5}$                     & 3              & $10^{-3}$ & 0.18    & 100  & 5.8  $\times 10^7$     & 0.198& -4.83$\times10^{-3}$& 120& 108&36& 0.58         \\
		S5        & 5.0    $\times 10^{-5}$                    & 3              & $10^{-3}$ & 0.36    & 100  & 5.8 $\times 10^7$     & 0.183& -6.42$\times10^{-3}$& 109& 63.3&36& 0.08         \\
		S6        & 1.0     $\times 10^{-4}$                    & 3              & $10^{-3}$ & 0.72    & 100  & 5.8 $\times 10^7$    & 0.168& -8.23$\times10^{-3}$& 96.1&38.6&24& 0.74         \\
		S7        & 2.0   $\times 10^{-4}$                       & 3              & $10^{-3}$ & 1.44    & 100  & 5.8 $\times 10^7$   & 0.159& -10.2$\times10^{-3}$& 82.0&26.3&24& -0.01         \\
		&                           &                &           &         &      &                    &&&&&&\\
		T1        & 1.25 $\times 10^{-5}$                       & 1.5            & $10^{-3}$ & 0.09    & 50   & 5.8 $\times 10^7$    & 0.222& -2.16$\times10^{-3}$& 176& 341 &48&-0.10          \\
		T3        & 1.25 $\times 10^{-5}$                       & 6              & $10^{-3}$ & 0.09    & 201  & 5.8 $\times 10^7$    & 0.196& -5.10$\times10^{-3}$& 101& 98.5&60& 0.83         \\
		T4        & 1.25 $\times 10^{-5}$                       & 12             & $10^{-3}$ & 0.09    & 402  & 5.8  $\times 10^7$   & 0.172& -7.23$\times10^{-3}$& 74.2&46.9&84& 0.53         \\
		&                           &                &           &         &      &                    &&&&&&\\
		Z1        & 1.25   $\times 10^{-5}$                     & 3              & $10^{-4}$ & 0.09    & 100  & 5.8 $\times 10^8$    & 0.187& -2.85$\times10^{-3}$& 115& 154&60&  0.40        \\
		Z3        & 1.25 $\times 10^{-5}$                       & 3              & $10^{-1}$ & 0.09    & 100  & 5.8 $\times 10^5$    & 0.278& -5.02$\times10^{-3}$& 185& 287&60&  -0.41        \\
		&                           &                &           &         &      &                    &&&&&&\\
	
		GABLS1    & 0                         &             & $10^{-1}$ & 0       & -    & 5.8 $\times 10^8$       				 & 0.263& -10.9$\times10^{-3}$& 205& 112 & 9& -\\
		
		WSBL-B24 & 4.0   $\times 10^{-5}$                      & 25 & $10^{-3}$ & 0.29    & 626  & 8.6 $\times 10^7$     &0.183 & $-25.6\times 10^{-3}$& 52.2 & 14.3 & 32 &  -0.31       \\
		&                           &                &           &         &      &                    &&&&&&\\
		WSBL-L19    & 4.0   $\times 10^{-5}$  & 25    & $10^{-3}$ & 0.29       & 626    &  8.6 $\times 10^7$ & 0.184      & $-24.6\times 10^{-3}$ & 51.9 & 15.6  & 30 & -\\
		VSBL-L19    & 4.0   $\times 10^{-5}$  & 25    & $10^{-3}$ & 0.29       & 2145   &  2.5 $\times 10^7$ & 0.0356     & $-3.08\times 10^{-3}$ & 5.93 & 0.915 & 23 &-\\
		\hline             
	\end{tabular}
\end{sidewaystable}

\subsubsection{Initialization and Time Development}
\label{subsubsec:initialization}
All simulations are initialized with a uniform velocity $u = G$ in the $x-$direction, with random perturbation of $0.1G$ added to the velocity components below $h_p=100$ m to trigger turbulence. The initial temperature profile increases linearly from $\theta_s$ to $\theta_\infty = \theta_s+\mathrm{\Delta}\theta$ at $h_i=200$ m and is equal to $\theta_\infty$ above. In preliminary simulations, initial conditions were varied by e.g. changing the initial inversion height $h_i$ or starting from a constant temperature $\theta_\infty$ and gradually cooling the surface temperature towards the final value $\theta_s$ (similar to the DNSs of \cite{Gohari2018} or LESs of L19 and \cite{Maronga2020}). Crucially, the initial conditions did not have an effect on the mean profiles in the final equilibrium state, except the time needed to reach this steady state. Furthermore, in the latter scenario, a notably stronger inertial oscillation in the upper part of the domain was observed. This can be attributed to turbulent perturbations penetrating deeply into the neutral free atmosphere, leading to differences between the horizontal velocity components and geostrophic wind, consequently triggering the inertial oscillation \citep{Blackadar1957, VandeWiel2010a}. By applying a temperature gradient up to a sufficient height from the onset of the simulations, there is a stably stratified layer above the initial turbulent perturbations, that prevents turbulence to enter into the neutral free atmosphere. 

In all  simulations, the amplitude of the inertial oscillation, quantified by considering the time development of $\langle u\rangle$ and $\langle v\rangle $ averaged over the upper 25\% of the domain, was smaller than 1\% of the geostrophic wind $G$. Therefore, we were not required to use forced damping methods in the upper part of the domain, such as the geostrophic damping approach proposed by \cite{Stipa2023} or the widely used Rayleigh Damping Layer \citep{Klemp1977, Lanzilao2022}. The latter is commonly applied in numerical simulations to prevent spurious reflection of gravity waves at the upper boundary, but unnecessary here due to the unstratified free atmosphere.  

To assess whether a simulation has reached thermal equilibrium, the instantaneous sum of the surface heat flux and the domain-integrated subsidence heating are diagnosed as:
\begin{equation}
	Q_{\mathrm{sum}}(t) = q_0(t) + \int_{0}^{L_z} \mathrm{\Gamma}_w z \frac{\partial \langle \theta \rangle}{\partial z }(t)\mathrm{d}z,
\end{equation}
which should be close to 0 in the steady state. The total simulation time for each case is reported in Table \ref{tab:overview}, and all statistics are collected over the last $T_{av}=2\pi/f_c \approx 12.56$ hours, corresponding to one inertial period. The ratio $R_q = Q_{\mathrm{sum}}/q_0$, averaged over $T_{av}$, was ensured to be smaller than 1\%, as reported in Table \ref{tab:overview}. In the remainder of the text, all variables can be assumed to be time-averaged and in the statistically steady state. 

\subsubsection{Resolution and Domain Size}
All simulations are performed on a domain of $L_x \times L_y \times L_z = 400^3$~m, equal to the GABLS1 configuration. The maximum boundary-layer depth $h$ of approximately 200~m in our cases is also comparable, although it is smaller in most cases (see next section and Table \ref{tab:overview}). To assess the sensitivity to domain size, two simulations of case S3T2Z2 were run, where the respective horizontal and vertical dimensions were doubled, which had negligible effect on mean profiles of first and second order statistics. The only observable difference was in the peak of the turbulent temperature fluctuation $\langle {\theta'^2}\rangle$ that occurs at the height of the inversion layer (not shown), for the case where the horizontal domain size was doubled. This peak in temperature variance is assumed to be related to internal waves in the very stable capping inversion \citep{Saiki2000, Lloyd2022}. As already argued by \cite{Beare2006}, domain lengths of multiple kilometres would be required to catch the full extent of these gravity waves, which is beyond the interest of the present work.

Regarding resolution, a grid size of $128 \times 128 \times 256$ points is utilized, corresponding to a horizontal grid spacing of $\mathrm{\Delta}_x = \mathrm{\Delta}_y = 3.125$ m and $\mathrm{\Delta}_z = 1.56$ m in the vertical direction. \cite{Allaerts2016} is followed by using a finer resolution in the vertical direction than in the horizontal, as is common for codes with pseudo-spectral discretization in horizontal directions \citep[e.g.][]{Stoll2009, Huang2013}. 

Because increasing stratification  leads to a decrease in TKE, the dependency on the grid resolution and SGS model is amplified in simulations where the stability is stronger \citep{Sullivan2016a, VanderLinden2019}. Therefore, the grid sensitivity is evaluated for the three scenarios exhibiting the strongest stratification, as measured by stability parameter $h_\theta/L_O$, with $L_O = \theta_r u_*^3 (\kappa g q_0)^{-1}$ the well-known Obukhov length  (cases S6, S7 and T4, see Table \ref{tab:overview}). Details on this resolution study are provided in Appendix 1, where it is concluded that the mesh is sufficiently fine for the purpose of this study. Specifically, mean profiles are shown to be hardly affected by doubling of the grid resolution, and surface fluxes change less than 2.5\%.

\subsubsection{Two Reference Simulations}
In addition to the simulations discussed above, two reference simulations are included which provide the ability to compare results to well-known cases from literature. First, the GABLS1 case was run, as was already used previously to validate the SP-Wind code for SBLs (see \cite{Allaerts2016} for more details). Note that the same grid spacing was used here as in the other simulations.
Secondly, to validate our code against a previous simulation that included subsidence, we ran the `Weakly Stable Boundary Layer (WSBL)' case from L19. Both our GABLS1 and `WSBL-B24' simulations are also included in Table \ref{tab:overview}. Additionally, the external parameters and bulk quantities from the WSBL and VSBL (Very Stable Boundary Layer) cases from L19 are reported in the last two rows. Of importance, cases from L19 employ a substantially larger global temperature difference of 25 K, while the subsidence rate lies in between our cases S4 and S5. The only difference between the WSBL and VSBL case is the geostrophic wind speed, which is reduced from 12 m/s to 3.5 m/s, thereby effectively increasing the Buoyancy number and lowering the surface Rossby number. More details on the WSBL-B24 and WSBL-L19 simulations, along with a comparison of mean profiles, is provided in Appendix 2. In general, the comparison yielded satisfactory results, especially when taking into account that the MicroHH code used by L19 employs a different discretization method and SGS model.

\section{Results}
\label{sec:results}

\subsection{Overview of Mean Profiles}
\begin{figure}[h]
	\includegraphics[width=\textwidth]{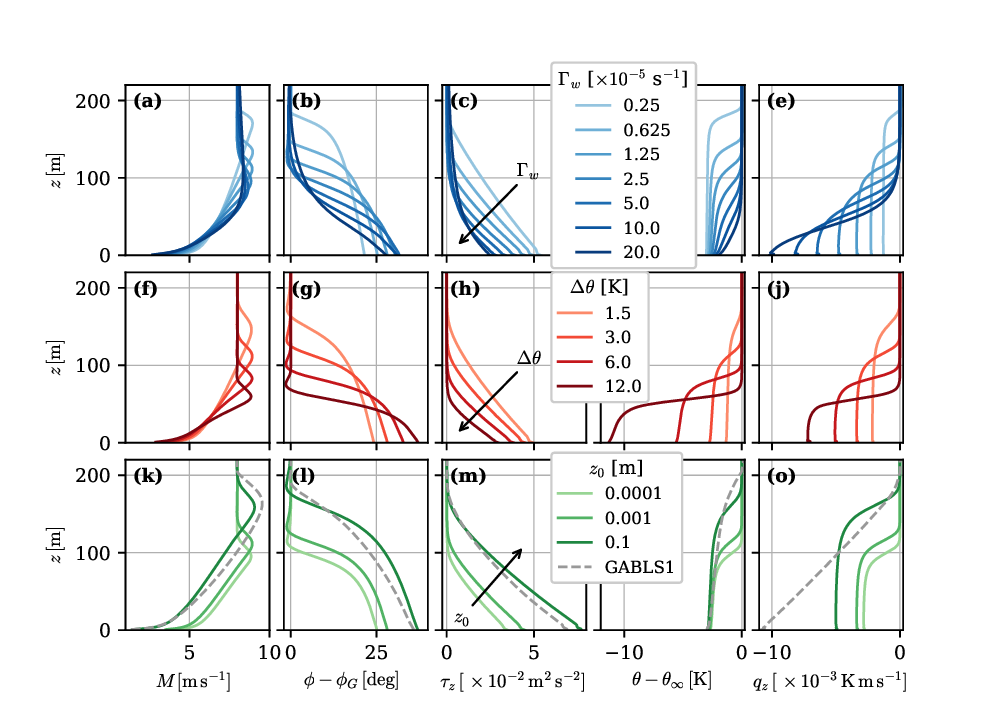}
	\caption{Overview of mean profiles. Columns represent horizontal wind speed magnitude, flow angle, vertical momentum flux, potential temperature and heat flux. Rows represent simulation set S, where subsidence rate is varied (top), set T, where the temperature difference is varied (middle), and set Z, where the surface roughness is varied (bottom). Lines with darker colors indicate higher values of $\mathrm{\Gamma}_w$, $\mathrm{\Delta}\theta$ and $z_0$, respectively. See table \ref{tab:overview} for a full description of the depicted simulations}\label{fig:profiles}
\end{figure}

The effect of each of the governing parameters in Eq. \eqref{eq:parameters} is explored by discussing the horizontal- and time-averaged profiles of momentum, temperature and their respective fluxes for sets S, T and Z as displayed in Fig. \ref{fig:profiles}. In all cases, the velocity magnitude $M=( \langle u\rangle^2 +  \langle v\rangle^2 )^{1/2}$ shows a supergeostrophic wind speed of about 9 m/s. This so-called low-level jet (LLJ) is often observed in both simulations and observations of the (stably) stratified atmosphere, and can have many possible causes \citep{Stull1988, Mahrt1999}. The formation of the LLJ in idealized LES of the SBL is often related to the inertial oscillation due to the Coriolis force. However, the LLJ in the present simulations is stationary. From the top row, it is evident that increasing subsidence rate decreases the height of the LLJ and the near-surface shear, whereas wind veer is increased. In addition, stronger subsidence reduces the turbulent vertical momentum flux $\tau_z = (\langle \tau_{xz}\rangle^2 + \langle \tau_{yz}\rangle ^2)^{1/2}$ (with $\tau_{iz} =   {u_i'w'} + \tau_{iz}^{sgs}$, i.e. the sum of resolved and subgrid shear stress) and increases the surface heat flux $q_0$. For low subsidence rates, there is a well-mixed layer with almost constant temperature capped by a strong inversion layer, resembling the widely studied `Conventionally Neutral Boundary Layer' (CNBL) profile with zero free-atmosphere stratification \citep[and references therein]{Liu2021b}. As subsidence rate increases, the height of this capping inversion decreases, and eventually the temperature gradient becomes more or less constant throughout the boundary layer for the cases with strongest subsidence. 

The second row of Fig. \ref{fig:profiles} shows the mean profiles for set T, where the temperature difference $\mathrm{\Delta}\theta$ between the free-stream and the surface is varied (see panel i). Effects of increasing $\mathrm{\Delta}\theta$ on the profiles of momentum are qualitatively similar to rising subsidence rate, i.e. the LLJ height descends, wind turns more sharply and surface stress decreases. Regarding the thermal profiles (panels i and j), the height of the inversion layer decreases while the surface heat flux increases. 

The effect of surface roughness (or Rossby number) on mean profiles is presented in the bottom row of Fig. \ref{fig:profiles}. It is evident that a rougher surface enhances turbulence generation and therefore increases vertical fluxes of momentum and heat, and the height of the LLJ. The temperature profiles seem unaltered in the lower part of the boundary layer, whereas only the height of the inversion layer increases with larger $z_0$. The grey dashed line in panels k-o, representing the GABLS1 case, displays roughly similar profiles to case C4, which has equal $z_0$. Note that for the temperature profile, $\theta_\infty = 266$ K is estimated to match the top of the inversion layer in the GABLS1 case with our simulations with subsidence (see also \ref{fig:scaled_profiles}). The main deviation is in the vertical heat flux $q_z$, which is approximately linear in the GABLS1 case, in agreement with local scaling \citep{Nieuwstadt1984}, while in our case the equilibrium equation \eqref{eq:balance} is satisfied (as further discussed in Sect. \ref{subsec:mean_profile_scaling}).

\subsection{Stable Boundary-Layer Depth}
\label{subsec:sbl_depth}
\begin{figure}[h]
	\centering\includegraphics[width=\textwidth]{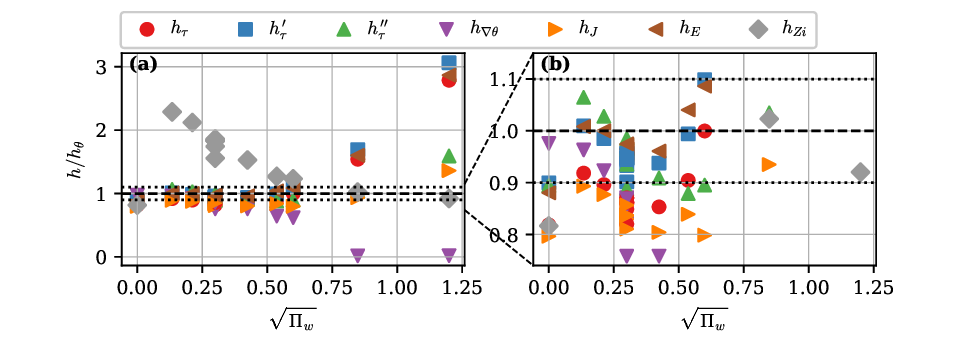}
	\caption{Comparison of SBL depth estimations calculated through the methods \ref{method1}-\ref{method6} explained in the text, normalized with the thermal boundary-layer height $h_\theta$. Grey diamonds refer to the estimated SBL depth $h_{Zi}$ obtained through equation \eqref{eq:h_theory}. Panel (b) is just a zoom of panel (a), while the black dotted lines indicate a 10\% deviation between the $h_\theta$ and the other SBL height estimations }\label{fig:height_ratios}
\end{figure}

In order to scale the vertical coordinate of the mean profiles in Fig. \ref{fig:profiles}, a length scale that characterizes the SBL depth is required. As the SBL often blends smoothly into the residual layer above, accurately measuring its height is difficult, resulting in the existence of numerous estimation approaches \citep[see e.g. p. 503]{Stull1988}. To illustrate this, six methods, frequently used in LES studies, are enlisted to characterize the SBL depth (partly inspired by \cite{Chinita2022}):
\begin{enumerate}
	\item $h_\tau$: the height where the momentum flux vanishes, estimated as the height where $\tau_z$ is 5\% of its surface value and then performing a linear extrapolation to the height where it would be 0, i.e. $h_\tau=z(\tau_z = 0.05u_*^2)/0.95$ \citep[e.g.][]{Beare2006, Huang2013, VanderLinden2019}. 
	\label{method1}
	\item $h_\tau'=z(\tau_z = 0.05u_*^2)/(1-0.05^{2/3})\approx 1.1h_\tau$: similar to $h_\tau$ but with different extrapolation \citep{Liu2021b}, motivated by the expected power-law shape of the shear stress profile in an SBL as derived by \cite{Nieuwstadt1984}, i.e. $\tau_z/u_* ^2 = (1-z/h)^{3/2}$.
	\item $h_\tau''$: obtained by performing a least squares fit of the power-law profile $\tau_z/u_* ^2 = (1-z/h)^{3/2}$ to the shear-stress profiles in the LES \citep{Narasimhan2023}.
	\item $h_{J}$: the position of the low-level jet (where velocity magnitude $M$ is a maximum.)
	\item $h_{\nabla\theta}$: the level where the temperature gradient is a maximum \citep{Sullivan2016a, McWilliams2023}.
	\item $h_{E}$: the height where the TKE reduces to 5\% of the value at the first model level \citep{Chinita2022}.
	\label{method6}
\end{enumerate}
In the present study however, the thermal boundary-layer height $h_\theta$ is used, because of its key role in the equations for thermal equilibrium (see Sect. \ref{subsec:thermal_equilibrium}). In practice, $h_\theta$ is approximated as the height where $(\langle \theta\rangle-\theta_s)/\mathrm{\Delta}\theta = 99\%$. The six enlisted SBL depth estimations are compared to $h_\theta$ in Fig. \ref{fig:height_ratios}. The figure shows $h_i/h_\theta$ for all cases. The choice of $\mathrm{\Pi}_w^{1/2}$ on the horizontal emphasizes that the largest difference between all estimation methods occurs for the two cases with strongest subsidence. Ignoring these two cases, the estimates based on turbulent shear stress (in particular $h_\tau'$ and $h_\tau''$) and TKE give similar results, and differ mostly less than 10\% from $h_\theta$ (see zoom in panel b.). Further, the height of the LLJ is consistently about 80-90\% of $h_\theta$. The level where the maximal temperature gradient occurs relative to $h_\theta$ descends with increasing subsidence rate. This estimation method, proposed by \cite{Sullivan2016a} to ensure the turbulent region above the LLJ is also included in the SBL depth, works well if there is a clear inversion layer with large gradients compared to the lower SBL, but fails in cases S6 and S7 since the maximum temperature gradient occurs very close to the surface. Regarding these two extreme cases, it is evident that estimates based on turbulent shear stress or TKE are much higher than the thermal boundary layer (see also Fig. \ref{fig:scaled_profiles} and related discussion). 

The markers at $\Pi_w = 0$ in Fig. \ref{fig:height_ratios} represent the GABLS1 case. Here, the thermal boundary-layer height is estimated to be the level of maximum curvature in the temperature profile, i.e. where $|\partial^2 \langle \theta\rangle/\partial z^2 |$ has a maximum. This results in $h_\theta=205$ m, which visually corresponds well to the transition from the inversion layer to the free atmosphere. Note that $\theta (z=h_\theta) \approx \theta_\infty \approx 266$ K, consistent with the value used in Fig. \ref{fig:profiles} and mentioned in the previous section. The location of the maximum temperature gradient is close to $h_\theta$, and considerably higher than $h_\tau$ and $h_J$. The latter two heights are nearly equal, and lower than the SBL depth estimation based on TKE, perfectly in line with previous studies \citep{Huang2013, Sullivan2016a, Chinita2022}.

Based on LES and field observations, semi-empirical formulas have been proposed to calculate the SBL depth from other flow characteristics \citep{Zilitinkevich2002, Zilitinkevich2007}. In order to assess whether these also work for the equilibrium SBL with subsidence, we consider the following well-established expression in dimensional form \citep{Zilitinkevich2007}: 
\begin{equation}
	\frac{1}{h_{\mathit{Zi}}^2} = \left( \frac{f_c}{ C_{\mathit{TN}} u_*} \right)^2 + \frac{f_c}{C_{\mathit{NS}}^2 \kappa u_* L_O} + \frac{f_c N_\infty}{\left(C_{\mathit{CN}} u_*\right)^2}
	\label{eq:h_theory}
\end{equation}
with $\kappa=0.4$ the Von Karman constant and $N_\infty = (\mathrm{\Gamma}_\theta g/\theta_r)^{1/2}$ the Brunt-Väisälä frequency in the free atmosphere, where $\mathrm{\Gamma}_\theta$ denotes the temperature gradient. For the empirical constants, we use $(C_{\mathit{TN}}, C_{\mathit{NS}}, C_{\mathit{CN}}) = (0.5, 0.78, 1.6)$ as determined by \cite{Liu2021a} and \cite{Narasimhan2023} from LES datasets. Equation \eqref{eq:h_theory} is essentially a squared reciprocal interpolation between the length scales that are relevant in a truly neutral boundary layer, SBL and CNBL, respectively \citep{Zilitinkevich2007}. Note that in the present simulations with subsidence, the CNBL contribution vanishes since $N_\infty=0$. Estimates based on Eq. \eqref{eq:h_theory} are represented by the grey diamonds in Fig. \ref{fig:height_ratios}, showing that the estimates are far off for the present SBLs with subsidence. 

\cite{Zili2002} proposed a correction to $h_{Zi}$ derived from a relaxation equation of the actual SBL height to account for synoptic scale vertical motions: $h_{\mathit{Zi}-\mathrm{cor}}= h_{\mathit{Zi}} + w_h t_R$, with $w_h$ the large-scale vertical velocity at the top of the SBL ($w_h = -\mathrm{\Gamma}_w h$ in our case). For the relaxation timescale $t_R$, \cite{Zilitinkevich2002} use $t_R = C_E /f_c$, while \cite{Zilitinkevich2007} recommend $t_R = C_E h/u_*$, with $C_E \approx 1$. Figure \ref{fig:height_ratios} however clearly shows that the error in $h_{Zi}$ is not linear in $\mathrm{\Gamma}_w$, implying that such a correction does not work here. 

\begin{figure}[h]
	\centering\includegraphics{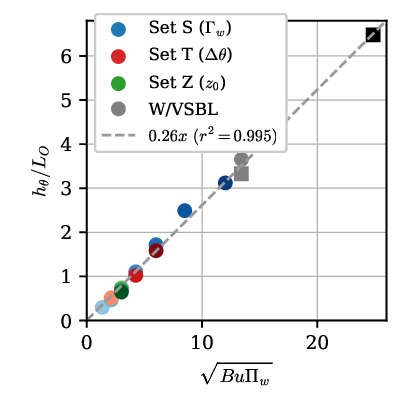}
	\caption{Stability parameter $h_\theta/L_O$ compared against the square root of product of non-dimensional Buoyancy and Subsidence numbers. Each point represents one simulation: blue circles indicate simulations of set S with varying subsidence rate, red circles represent simulations of set T with varying temperature difference, and green circles are simulations of set Z with varying surface roughness. Darker colors imply higher values. The grey circle indicates WSBL-B24 case, while the grey and black squares represent data from the WSBL and VSBL cases of L19. The grey dashed line is the linear least-squares fit to the data: $y=0.26x$ (cf. Eq. \eqref{eq:h/L}) with coefficient of determination $r^2=0.995$ }\label{fig:hL_correlation}
\end{figure}

Based on our simulations, a different scaling for thermal boundary-layer height is proposed:
\begin{equation}
	\frac{h_\theta}{L_O} = \frac{h_\theta \kappa g q_0}{\theta_r u_*^3} \approx C_h \sqrt{\mathrm{\Pi}_w \mathit{Bu}} ,
\label{eq:h/L}
\end{equation}
with $L_O$ the Obukhov length, and $\mathit{Bu}$ and $\mathrm{\Pi}_w$ the non-dimensional groups introduced before (Eq. \ref{eq:parameters}). The ratio $h/L_O$ is typically referred to as the stability parameter \citep[e.g.][]{Holtslag1986, Heisel2023}. 
$C_h \approx 0.26$ is determined through a least-squares fit with coefficient of determination $r^2=0.995$, as shown in Fig. \ref{fig:hL_correlation}. To assess the validity of Eq. \eqref{eq:h/L} in a wider range of simulations, the WSBL and VSBL cases from L19 are also included in Fig. \ref{fig:hL_correlation}. Since $h_\theta$ is not reported, $h_\theta \approx 1.1h_\tau$ is estimated based on our own WSBL simulation (see also Table \ref{tab:overview} and Fig. \ref{fig:height_ratios}). Of note, Eq. \eqref{eq:h/L} does not depend on the third external parameter ($z_0$ or $\mathit{Ro_0}$), because including $\log(\mathit{Ro_0})$ resulted in only negligible changes of $r^2$. In Sect. \ref{subsec:GDL}, the relation in Eq. \eqref{eq:h/L} is used to estimate $h_\theta$ from external parameters only.

\subsection{Scaling of Steady Mean Profiles}
\label{subsec:mean_profile_scaling}
\begin{figure}[h]
	\centering\includegraphics[width=\textwidth]{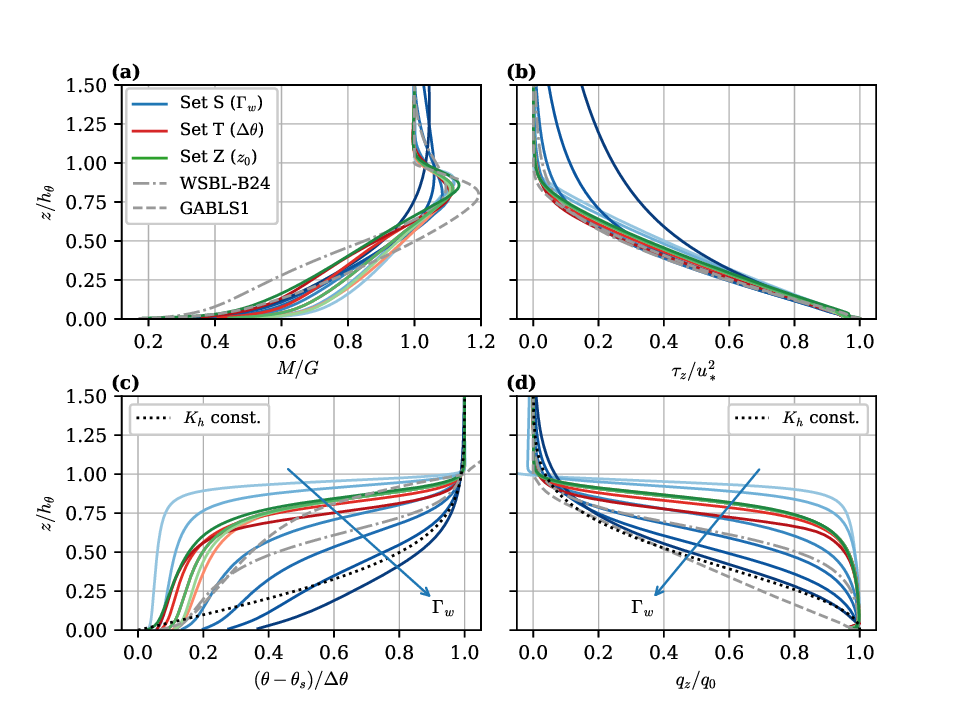}
	\caption{Mean profiles of (a): horizontal wind magntiude scaled with geostrophic wind speed, (b) momentum flux normalized with surface momentum flux, (c) potential temperature scaled with global temperature difference and (d) heat flux scaled with surface heat flux. The vertical coordinate is scaled with thermal boundary-layer height $h_\theta$. Line colors are the equal to Fig. \ref{fig:profiles}: blue lines indicate varying subsidence rate (set S), red lines represent varying temperature difference (set T) and green lines denote varying surface roughness (set Z). The grey dashed and dash-dotted lines represent the GABLS1 and WSBL-B24 cases, respectively. In the bottom row, the black dotted lines are obtained analytically for constant eddy-diffusivity (see Eqs. \eqref{eq:analytical_profiles}), while the blue arrows indicate direction of increasing subsidence rate }\label{fig:scaled_profiles}
\end{figure}

In order to assess similarity of mean variables, Fig. \ref{fig:scaled_profiles} presents scaled profiles of the simulations reported in Table~\ref{tab:overview}. The horizontal velocity magnitude is scaled with the geostrophic wind $G$ (Fig. \ref{fig:scaled_profiles}a), temperature is normalized with the global temperature difference $\mathrm{\Delta}\theta$ (Fig. \ref{fig:scaled_profiles}c), while the vertical momentum- and heat fluxes are normalized with their respective surface values ( Figs. \ref{fig:scaled_profiles}b, d). The thermal boundary-layer height $h_\theta$ is used to normalize the vertical coordinate. 
Panel a) shows not only that the height where the LLJ occurs coincides in most cases (cf. Fig. \ref{fig:height_ratios}), but also that the maximum wind speed is roughly equal ($\sim 1.1G$). The maximum wind speed is slightly lower for the cases with strongest subsidence (darker blue lines), while the LLJ is slightly stronger for the GABLS1 case ($\sim 1.2G$, dashed grey line). The scaled momentum flux profiles in Fig. \ref{fig:scaled_profiles}b coincide remarkably well, even with the GABLS1 case. 

In line with the observations in the preceding section, only the the cases with strongest subsidence rate clearly deviate from the other profiles, i.e. the height where the shear stress vanishes is significantly higher than the depth of the thermal boundary layer. Since the temperature gradient above $h_\theta$ is negligible, this `decoupling' of the thermal and turbulent boundary layer implies that a large portion of the turbulent motions is not damped by stratification. Figure \ref{fig:resolution_sensitivity}e in Appendix 1 shows that in these cases, the gradient Richardson number in the thermal boundary layer does not reach the critical value of $~0.25$, which offers an explanation for the survival of turbulence and its penetration high into the domain \citep[e.g.]{Sullivan2016a}.

Figures \ref{fig:scaled_profiles}c and d reveal that the shape of the normalized temperature and heat flux profiles is mainly determined by the subsidence rate $\mathrm{\Gamma}_w$, as the blue curves clearly do not overlap. By contrast, the green and red curves, which indicate that the global temperature difference (or $\mathit{Bu}$) and surface roughness (or $\mathit{Ro_0}$) are varied, appear to be very similar. This finding is corroborated by L19, who remark that the temperature profiles in their WSBL and VSBL cases (where $\mathrm{\Pi}_w$ remains equal while both $\mathit{Bu}$ and $\mathit{Ro_0}$ are changed, see Table \ref{tab:overview}) have the same overall shape. As discussed in the previous section, weak subsidence results in a nearly neutral boundary layer with strong capping inversion, whereas high subsidence rates result in a seemingly exponential (convex) temperature profile. The former `convex-concave-convex' shape was also observed by L19, while the field observations, where they compared their simulations to, showed more convex or linear temperature profiles. They attributed this discrepancy to the lack of a radiation model or erroneous subsidence profile and magnitude. The present results demonstrate that a higher subsidence rate indeed leads to a more convex or linear temperature profile. 

\begin{figure}[h]
	\centering
	\includegraphics{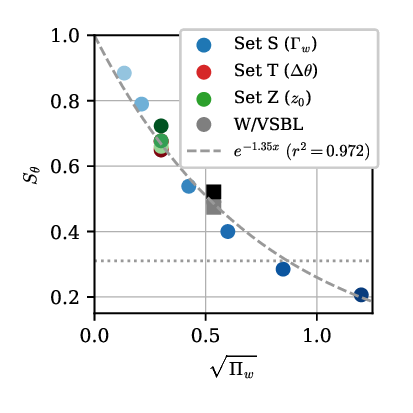}
	\caption{Dependency of the shape factor of the temperature profile $S_\theta$ (Eq. \eqref{eq:shape_factor}) on square-root of the non-dimensional subsidence rate $\mathrm{\Pi}_w$. Colors and markers are equal to Fig. \ref{fig:hL_correlation}. The grey dashed line indicates the least-squares fit $y=\exp(-C_S x)$ with $C_S=1.35$ and coefficient of determination $r^2=0.972$. Horizontal dotted line indicates the shape factor that would be obtained for constant eddy-diffusivity (see Eq. \eqref{eq:analytical_profiles})}\label{fig:S_correlation}
\end{figure}

To quantify the shape of the temperature profile, the shape factor $S_\theta$ as defined in Eq. \eqref{eq:shape_factor} is used. 
Figure \ref{fig:S_correlation} displays $S_\theta$ for all simulations as a function of $\mathrm{\Pi}_w^{1/2}$. The good collapse of all cases from simulation sets T and Z (green and red markers) re-emphasizes that the scaled temperature profile is mainly determined by the dimensionless subsidence rate $\mathrm{\Pi}_w = \mathrm{\Gamma}_w/f_c$. For small subsidence rates, the shape factor approaches 1, as the scaled temperature profile has a rectangular shape with a thin inversion layer capping the near-neutral boundary layer (see Fig. \ref{fig:scaled_profiles}c). For $\mathrm{\Pi}_w \rightarrow 0$ and $S_\theta \rightarrow 1$, the temperature gradient in the inversion layer would go to infinity. Therefore, there should be a physical upper limit on $S_\theta$ and $\mathrm{\Pi}_w$ for which equilibrium is still possible, which it is expected to depend on a critical Richardson number in the inversion layer. Exploring this limit with LES however would require very fine grid resolutions in the strongly stratified inversion layer, which is beyond the scope of the present work (see also Appendix 1). On the other hand, increasing subsidence rate reduces the shape factor, which seems to go asymptotically towards 0 as shown by Fig.~\ref{fig:S_correlation} (although it can be derived that in the limit of laminar flow, $S_\theta \approx 0.31$ (see below), as indicated by the horizontal dotted line). 

To describe the dependency of $S_\theta$ on the subsidence rate, a linear regression to fit the exponential function is used,
\begin{equation}
	S_\theta = \frac{q_0}{\mathrm{\Gamma}_w h_\theta \mathrm{\Delta}\theta} = \exp\left(-C_S (\mathrm{\Pi}_w)^{1/2}\right)
	\label{eq:S_correlation}
\end{equation}
to the data, resulting in $C_S = 1.35$. A least-squares fit of the curve $\exp\left(-C_S \mathrm{\Pi}_w\right)$ (i.e. without the square root) resulted in a poor match with the data ($r^2=0.68$), motivating the choice for $\mathrm{\Pi}_w^{1/2}$ on the horizontal axis of Fig. \ref{fig:S_correlation}.  The WSBL and VSBL cases from L19 (square markers) agree very well with the curve in Eq. \ref{eq:S_correlation}, even though they have a significantly larger temperature difference (and $\mathit{Bu}$) and different geostrophic winds than the other simulations. This reinforces the conclusion that the temperature shape factor is predominantly controlled by subsidence rate.

Considering the normalized heat flux profiles in Fig. \ref{fig:scaled_profiles}d, a similar statement can be made, i.e. the Buoyancy and Rossby number only have a small effect while subsidence rate mainly determines the shape of the profiles. This analogy is expected, since the equilibrium profiles of temperature and heat flux are directly related through the subsidence rate: 
\begin{equation}
	\frac{\partial \langle q_z \rangle }{\partial z} = \mathrm{\Gamma}_w z \frac{\partial \langle \theta \rangle }{\partial z}
	\label{eq:balance3}
\end{equation}
(see also Eq. \eqref{eq:balance2}). The equation above explains for example why the vertical gradient of the turbulent heat flux approaches zero near the surface, as well as the correlation between the strong temperature gradient and large change in turbulent heat flux in the inversion layer.

By contrast, the GABLS1 case (grey dashed line in Fig. \ref{fig:scaled_profiles}d), displays a linear heat flux profile. For such a quasi-stationary SBL without subsidence, \cite{Nieuwstadt1984} already derived analytical expressions for the vertical fluxes of heat and momentum. For the heat flux, a linear profile $q_z/q_0 = (1-z/h)$, was indeed obtained, in agreement with the LES results of the GABLS1 case. Assuming both the Richardson numbers based on fluxes and gradients to be equal to 0.2 throughout the boundary layer, \cite{Nieuwstadt1984}  derived $\tau_z/u_* ^2 = (1-z/h)^{3/2}$. Given that the GABLS1 profile in Fig. \ref{fig:scaled_profiles}b approximately follows this equation (e.g. \cite{Beare2006}), and most of the profiles from present simulations have a similar shape; Nieuwstadt's momentum flux profile describes even the SBL with subsidence reasonably well in most cases. The fact that the heat flux profiles do not match is a consequence of the additional subsidence term in the temperature equation (see Eqs. \eqref{eq:balance2} and \eqref{eq:balance3}), which makes solving the system of equations from \citet[][see his Appendix]{Nieuwstadt1984} less straightforward. Therefore, we did not succeed in deriving analytical flux profiles by adapting Nieuwstadt's derivation with an additional subsidence term. 

A different approach to finding analytical expressions for the temperature and heat flux profiles, would be to assume a gradient diffusion hypothesis: $q_z = -K_h \partial\langle\theta/\rangle\partial z$, with $K_h$ the (effective) eddy diffusivity. Substituting this in Eq. \eqref{eq:balance3}, leads to a linear differential equation for the potential temperature:
\begin{equation}
	\frac{\partial}{\partial z} \left(-K_h \frac{\partial  \langle\theta\rangle }{\partial z} \right)= \mathrm{\Gamma}_w z \frac{\partial \langle \theta\rangle  }{\partial z}
	\label{eq:balance_Kh}
\end{equation}
The most straightforward solution is obtained for constant $K_h$. With the appropriate boundary conditions, it can be shown that Eq. \eqref{eq:balance_Kh} then leads to the following normalized temperature and heat flux profiles: 
\begin{equation}
	\frac{\langle\theta\rangle - \theta _s}{\mathrm{\Delta}\theta} = \mathrm{erf}\left(\frac{{z}}{\sqrt{2}{\sigma_\theta}} \right), \quad \frac{q_z}{q_0} = \exp\left({-\frac{{z}^2}{2{\sigma_\theta}^2}} \right)
	\label{eq:analytical_profiles}
\end{equation}
where $\mathrm{erf}$ is the error function and $\sigma_\theta = (K_h/\mathrm{\Gamma}_w)^{1/2}$, a length scale composed of the effective eddy diffusivity and subsidence rate. We remark that this length scale resembles the more familiar Ekman depth scale $D_E = (2K_m/f_c)^{1/2}$, where the eddy diffusivity is replaced by the eddy viscosity, and the Coriolis frequency plays the role of subsidence rate. This scale can be derived directly from a momentum balance, see e.g. \cite{Zili2002}. Moreover, the turbulent eddy diffusivity $K_h$ in Eq. \ref{eq:balance_Kh} could be easily replaced by a (constant) molecular diffusivity for temperature, in which case the profiles in Eq. \eqref{eq:analytical_profiles} can be interpreted as a result of laminar diffusion in the presence of subsidence, without any turbulent transport. Accordingly, $\sigma_\theta$ then becomes analogous to the `viscous Ekman layer depth' that is often used in DNS of Ekman layers \citep{Coleman1990}.

In order to obtain the expressions in Eq. \eqref{eq:analytical_profiles} as a function of $z/h_\theta$, we can use that $(\langle\theta\rangle-\theta_s)/\mathrm{\Delta}\theta = 0.99$ at $z=h_\theta$ by definition, resulting in $\sigma_\theta/h_\theta = 0.388$. This allows to compare the non-dimensional temperature and heat flux profiles for constant $K_h$ to the LES results, as shown by the black dotted lines in Figs. \ref{fig:scaled_profiles}c and d. It is clear that the LES profiles do not match with these analytical solutions, which is due to the fact that the eddy diffusivity $K_h$ is a function of $z$ rather than constant. Remarkably, the heat flux profile in the case with the strongest subsidence (S7, darkest blue line) agrees rather well with the analytical profile. Inspection of the $K_h$-profiles (not shown) reveals that this is indeed the case where the eddy diffusivity shows the smallest variation with height. From the definition of the thermal shape factor (Eq. \eqref{eq:shape_factor}) and the expressions in Eq. \eqref{eq:analytical_profiles}, it can be shown that $S_\theta = 0.31$ for the case of constant $K_h$, as indicated by the horizontal dotted line in figure \ref{fig:S_correlation}. The fact that $S_\theta$ drops below this value for cases S6 and S7, suggests that there exists a local minimum for a certain value of $\mathrm{\Pi}_w$, since the limit of Eq. \eqref{eq:analytical_profiles} should be reached under extremely stable conditions where all turbulence is eliminated and the eddy diffusivity is replaced by the (constant) molecular thermal diffusivity.

The accuracy of the solution to Eq. \eqref{eq:balance_Kh} is expected to improve by including a height-dependent eddy diffusivity. Various parametrizations have been proposed for eddy viscosity \citep{Brost1978, Nieuwstadt1983}, which could be adapted to the eddy diffusivity through $K_h \approx K_m/\mathit{Pr_t}$, with $\mathit{Pr_t}$ the turbulent Prandtl number that may or may not be height dependent. Due to the large number of assumptions there, we did not find a general parametrization that described the actual $K_h(z)$ in the LESs accurately enough to produce solutions to Eq. \eqref{eq:balance_Kh} that matched the LES results satisfactory well. Hence, finding analytical solutions for equilibrium SBL profiles with subsidence, similar to \cite{Narasimhan2023} for the case without subsidence, remains a topic for future work.

\subsection{A Simple Geostrophic Drag Law to Close the Model}
\label{subsec:GDL}
\begin{figure}[h]
	\centering
	\includegraphics{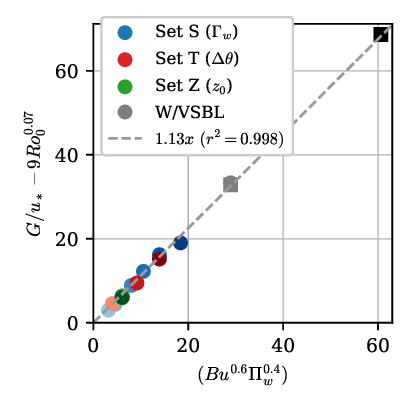}
	\caption{Correlation between the reduced friction coefficient and product of non-dimensional Buoyancy and Subsidence numbers, leading to the geostrophic drag law in Eq. \eqref{eq:GDL}. See figure \ref{fig:hL_correlation} for meaning of colors and markers. The grey dashed line indicates the least-squares linear fit $y=1.13x$ with coefficient of determination $r^2=0.998$ }\label{fig:GDL_correlation}
\end{figure}

From the empirical correlations \eqref{eq:h/L} and \eqref{eq:S_correlation} that were developed in the preceding section, the surface heat flux and thermal boundary-layer height can be predicted based on the external simulation parameters $\mathrm{\Pi}_w$ and $\mathit{Bu}$, if the friction velocity $u_*$ is provided. A so-called geostrophic drag law (GDL) can be used to determine the geostrophic drag coefficient $u_*/G$ \citep[e.g.][]{Zilitinkevich2002a}). A variety of GDLs has been proposed for different conditions, often in complicated formulations with dependencies on internal flow parameters such as boundary-layer height and geostrophic wind direction, and multiple empirical constants \cite[][most recently]{Liu2021, Narasimhan2023}. Instead, a formulation preferred is one where the dependent variables are fully determined by the relevant independent variables (given in Eq. \eqref{eq:parameters}), following \cite{Swinbank1974}. For a neutral ABL and without taking wind direction into account, the basic relation $u_*/G = 0.111 \mathit{Ro_0}^{-0.07}$ was found, valid in the range $10^4 <\mathit{Ro_0} < 10^9$. Starting from this expression, the deviation between the simulations here and this neutral case are calculated. Realizing that this departure must be proportional to stability and subsidence effects, and thus eventually finding excellent agreement for the power-law form $G/u_* - (0.111 \mathit{Ro_0}^{-0.07})^{-1} = C_g \mathit{Bu}^{p_1} \mathrm{\Pi}_w^{p_2}$. The suitable values for $p_1$ and $p_2$ were determined through trial and error to be approximately 0.6 and 0.4, respectively. Figure \ref{fig:GDL_correlation} shows that a least squares fit with $C_g=1.13$ works very well for the data in both the present study and L19. 

The friction velocity $u_*$ can then be predicted from solely external parameters via:
\begin{equation}
	u_* = \frac{G}{C_r \mathit{Ro_0}^{0.07} + C_g \mathit{Bu}^{0.6}\mathrm{\Pi}_w^{0.4}}.
	\label{eq:GDL}
\end{equation}
with $C_r = 1/0.111 = 9$ and $C_g = 1.13$. 
\begin{figure}[h]
	\centering
	\includegraphics[width=\textwidth]{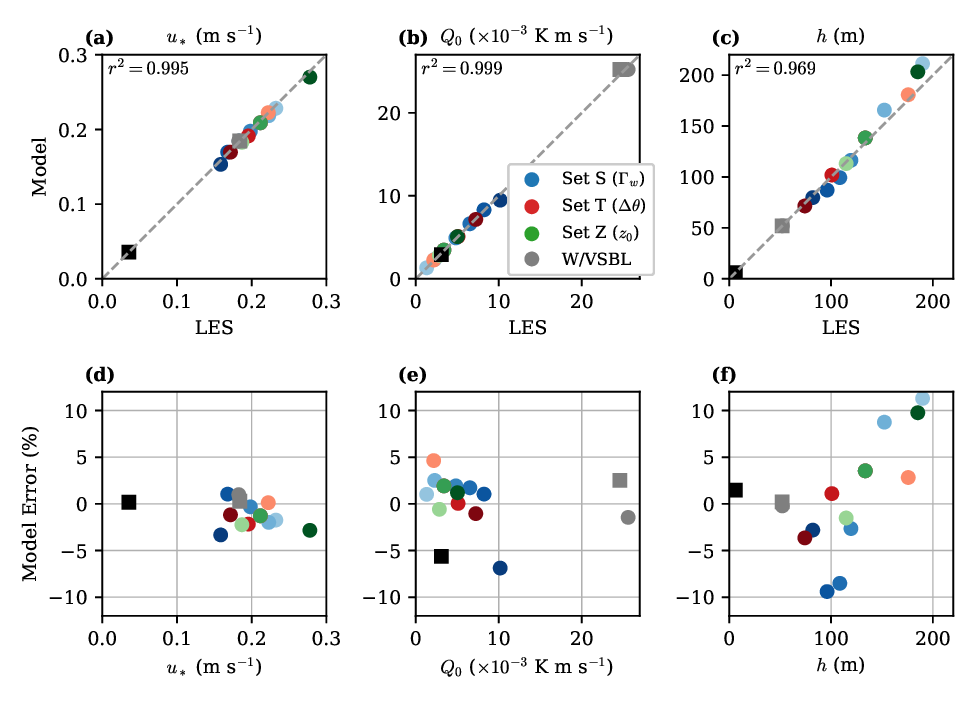}
	\caption{(a-c): estimations of friction velocity, surface heat flux and thermal boundary-layer height through model equations \eqref{eq:h/L}, \eqref{eq:S_correlation} and \eqref{eq:GDL} compared against values from LES. (d-f): deviation of model predictions from LES value in percent. The meaning of colors and symbols is explained in the caption of Fig. \ref{fig:hL_correlation}. The grey dashed lines in the top row indicate $y=x$, while the $r^2$ values denote the coefficient of determination}\label{fig:model}
\end{figure}

Together with Eqs. \eqref{eq:h/L} and \eqref{eq:S_correlation}, Eq. \eqref{eq:GDL} forms a closed model that can be used to estimate the friction velocity, surface heat flux and thermal boundary-layer height from external parameters in the present simulations of the SBL with subsidence. The model predictions are compared to the LES data in Fig. \ref{fig:model}, where the top row shows correlations plots and the bottom row the relative error of the estimation. Every marker represents one LES case, where the squares indicate the WSBL and VSBL cases of L19. Maximal errors below 5\% for $q_0$ and $u_*$, and around 10\% for $h_\theta$ are found. The match with the VSBL case from L19 (black squares in Fig. \ref{fig:model}) suggests that the model also covers the very stable boundary layer regime, which exhibits intermittent turbulent bursts resulting from unstable internal waves \citep{VanDerLinden2020}.

\begin{figure}[h]
	\centering
	\includegraphics[width=\textwidth]{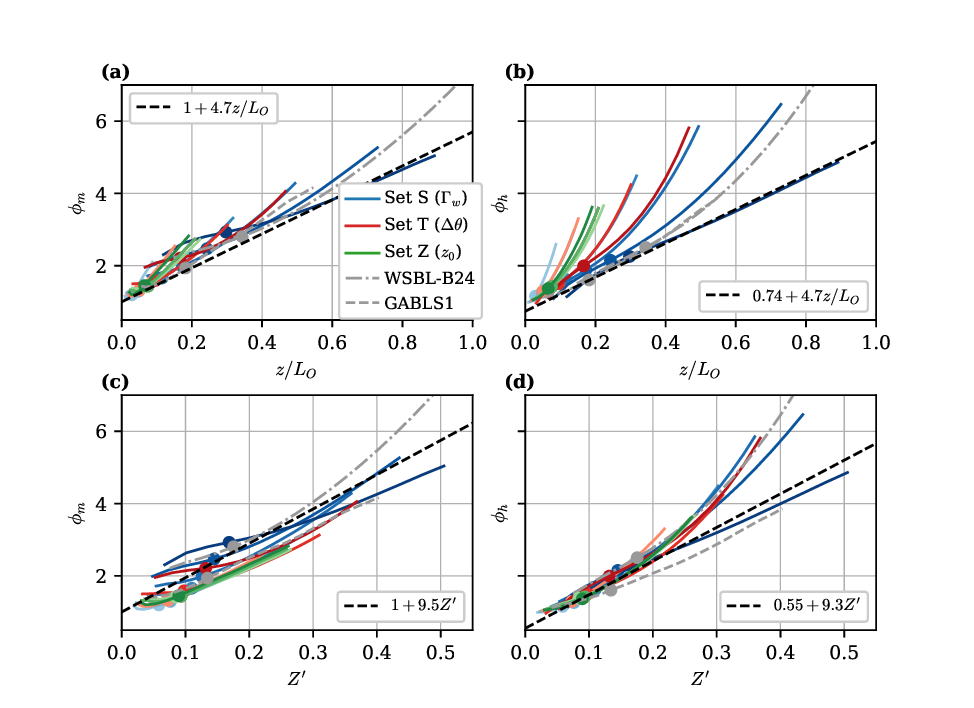}
	\caption{Comparison of classic Monin--Obukhov similarity theory (a,b) and the mixed scaling similarity proposed by \cite{Heisel2023} (c,d), by means of the dimensionless gradients for momentum (a,c) and temperature (b,d). The composite scaling parameter $Z'$ is defined in Eq. \ref{eq:Z'}. Only data in the range $0.03 < z/h_\theta < 0.3$ is included. The meaning of the colors corresponds to Fig. \ref{fig:scaled_profiles}, and the linear black dashed lines correspond to empirical relations from (a,b) \cite{Businger1971} and (c,d) \cite{Heisel2023} }\label{fig:MOST_surface}
\end{figure}

\subsection{Monin--Obukhov Similarity in the Presence of Subsidence}
\label{subsec:MOST}
In this final section, we compare the results of the LESs of the equilibrium SBLs with subsidence to MOST (see \cite{Foken2006} for a historical overview). Within this framework, the dimensionless gradients for momentum and heat play a central role:
\begin{equation}
	\phi_m = \frac{\kappa z}{u_*}\left[  \left(\frac{ \partial u }{\partial z}\right)^2 +  \left(\frac{\partial  v }{\partial z}\right)^2\right]^{1/2}, \quad \mathrm{and}
	\label{eq:phi_m}
\end{equation}
\begin{equation}
	\phi_h = \frac{\kappa z}{\theta_*} \frac{\partial  \theta }{\partial z},
	\label{eq:phi_h}
\end{equation}
where $\theta_* = -q_0/u_*$ is the friction temperature. 
MOST assumes that these non-dimensional gradients are universal functions of the ratio $z/L_O$ in the surface layer, and do not depend on other length scales such as surface roughness or ABL depth. Under weakly stable stratification ($0 < z/L_O \lesssim 1$), the consensus is that $\phi_{m,h}$ are linear functions of $z/L_O$ \citep[][and references therein]{Heisel2023}. Figures \ref{fig:MOST_surface}a and b show the dimensionless velocity and temperature gradients in the LESs with subsidence from the present study. Following \cite{Heisel2023}, data between $z=0.03h_\theta$ and $0.3h_\theta$ is included, while the traditional limit of the surface layer is $z\approx 0.1h$. This height is illustrated by circular markers in Fig. \ref{fig:MOST_surface}. The widely used empirical similarity relations of the form $\alpha_{m/h} + \beta_{m/h} (z/L_O)$ from \cite{Businger1971} are indicated by the black dashed lines. It is clear that close to the surface (for $z<0.1h_\theta$), the linear similarity relations hold reasonably well. One could argue that the match with MOST close to the surface is merely due to the fact that LES models used the flux-gradient relations from MOST to determine the surface fluxes, thereby enforcing this match. However, as discussed in Appendix 3, the heat flux due to subsidence at the first grid level is negligible compared to the surface heat flux, justifying the use of MOST as a boundary condition. For the dimensionless temperature gradient above $0.1h_\theta$, a departure from the linear profile depending on Buoyancy and subsidence numbers is observed. Remarkably, it seems that the simulations with stronger buoyancy and subsidence effects are closer to the linear profile, as case S7 even (darkest blue line) exactly collapses on the relations from \cite{Businger1971}.

In recent work by \cite{Heisel2023}, a new mixed scaling parameter for similarity in the SBL was introduced. Mixed scaling arguments have have been proposed successfully before, for instance the scaling of streamwise Reynolds stresses with a combination of the inner and outer velocity scales \citep{Degraaff2000}. 
In order to incorporate the effect of the boundary-layer depth $h$ in MOST, \cite{Heisel2023} propose to multiply the traditional scaling argument $z/L_O$ by $(h/L_O)^{-1/2}$. It is shown that this new parameter $Z=z/(L_Oh)^{1/2}$ provides improved similarity compared to the original MOST for high-resolution LES data from \cite{Sullivan2016a} and field measurements from the CASES-99 campaign \citep{Poulos2002}, especially in the strongly stable regime. In an effort to corroborate their conclusions with an independent dataset, we compare the mixed scaling approach to data from the present steady SBLs with subsidence. In Sect. \ref{subsec:sbl_depth}, $h_\theta/L_O$ is proportional to $(\mathit{Bu}\mathrm{\Pi}_w)^{1/2}$ (see Eq. \eqref{eq:h/L}), allowing the expression to be, 

\begin{equation}
	Z\approx Z' = \frac{z}{h_\theta} C_h^{1/2} (\mathit{Bu}\mathrm{\Pi}_w)^{1/4}
	\label{eq:Z'}
\end{equation}
thereby replacing one of the unknown internal parameters by external ones. Moreover, by cancelling $L_O$ in the scaling argument $Z$, the problem of self-correlation due to common divisors  $u_*, \theta_*$ in $\phi_{m/h}$ and $z/L_O$ is effectively eliminated (as discussed by e.g. \cite{Klipp2004, Baas2006, Sorbjan2010, Heisel2023}). Figures \ref{fig:MOST_surface}c and d show the dimensionless momentum and temperature gradients as function of $Z'$, along with the linear relations (black dashed lines) proposed by \cite{Heisel2023}. The collapse is considerably improved for the $\phi_h$ profiles compared to the traditional MOST, although $\phi_h(Z')$ departs from the linear relation for $Z' > 0.2$. The enhancement in the similarity for the dimensionless momentum gradient in Fig. \ref{fig:MOST_surface}c is less evident. It seems that the $\phi_m$ profiles for simulation sets T and Z match rather well, while the cases with strong subsidence from set S (dark blue lines) have a larger offset. 

A last point to underline here is that \cite{Nieuwstadt1984} introduced a `local scaling` framework in order to extend MOST to the entire SBL. Plots similar to Fig. \ref{fig:MOST_surface} with height-dependent values of the fluxes, and therefore a local Obukhov length, yield a substantially better collapse (not shown). For this local scaling system, the similarity relations from from \cite{Businger1971} with $\beta_{m/h}=4.7$ overestimate the slope, while $\beta_{m/h}=3$ as suggested previously by \cite{Chinita2022} fits better to the present LES data.

\section{Conclusions}
\label{sec:conclusions}
The {statistically steady} stably stratified ABL in the presence of subsidence has been investigated using a set of idealized large-eddy simulations, forced by a constant geostrophic wind $G=8$~m~s$^{-1}$ and a linearly increasing subsidence velocity profile. By dimensional analysis, we show that the problem is governed by three non-dimensional groups: a surface Rossby number, a Buoyancy number and a dimensionless subsidence rate. To investigate the impact of these external parameters on the SBL, a range of different surface roughness lengths, subsidence rates and surface temperatures is considered. All simulations reached a {truly steady} state after 12 to 70 hours, where the cooling heat flux at the surface is compensated by the heating due to subsidence, {resulting in thermal equilibrium}. Hence, the present set of LES experiments extends previous research that showed subsidence can lead to a steady SBL \citep{Mirocha2010, VanderLinden2019} to a wider and less extreme range of conditions. Moreover, it allows to systematically investigate the impact of each governing parameter separately.

An examination of mean profiles reveals that all cases exhibit a low-level jet and constant temperature above some level. For increasing subsidence rate, the SBL changes from a near-neutral well-mixed layer capped by a strong inversion to a shallow, more stable ABL without capping inversion. Increasing the global temperature difference (i.e. decreasing the surface temperature) increases surface heat flux and reduces the height of the inversion layer, while a higher surface roughness results in a deeper SBL with more turbulence.

The thermal boundary-layer height is a consistent characterization of the SBL depth in thermal equilibrium, and is compared to previously used definitions of boundary-layer height in the SBL. Scaling the mean profiles with this length scale and the surface values results in an agreeable similarity of turbulent shear stress, except in cases with very strong subsidence, where the turbulence penetrates deep into the neutral free atmosphere. The normalized profiles of temperature and heat flux collapse for varying surface roughness or temperature, as their shape is mainly determined by the subsidence rate. Analytical expressions for these scaled mean profiles are derived for constant turbulent eddy-diffusivity $K_h$, but a more accurate parametrization of $K_h$ would be required to describe the profiles of the present LES experiments.

From empirical correlations for the stability parameter $h_\theta/L_O$, thermal shape factor $S_\theta$ and a new unidirectional geostrophic drag law, a model is formed that estimates relevant internal flow properties based on the three governing non-dimensional groups. For the present set of LES data, complemented by a weakly and very stable boundary layer case from \cite{VanderLinden2019}, the errors in predictions of surface heat flux and friction velocity are shown to be smaller than 5\%, while the thermal boundary-layer height is estimated with a 10\% accuracy. 

A comparison of the dimensionless velocity and temperature gradients of the equilibrium profiles to MOST shows acceptable agreement in the surface layer. Higher up, similarity of non-dimensional temperature gradient is shown to be improved by using a new scaling argument that includes the thermal boundary-layer height and the non-dimensional Buoyancy and subsidence numbers, derived from the mixed scaling argument from \cite{Heisel2023} and the empirical correlation of the stability parameter.

With the present investigation, light is shed on the complex relationship between subsidence, SBL depth, and surface fluxes, as well as the internal distribution of mean profiles of temperature, momentum and their vertical fluxes. Although a truly steady state may be encountered rarely in reality, the present stationary simulations could also be interpreted as an Eulerian view of a continuous transition from e.g. weak to strong subsidence or small to large vertical temperature difference, with each simulation representing a stage of the transition \citep{Chung2012}.

Finally, the steady-state SBL with subsidence could serve as a new benchmark case for LES. Moreover, the same principle of subsidence heating that compensates for surface cooling could be used in DNS of the SBL. The resulting stationary state allows for long averaging times and more formal comparison of multiple experiments (with different grid resolution, SGS models, wind-farm layout etc.) than a transient simulation with strong inertial oscillation such as the GABLS1 case \citep{Sescu2014, Maronga2022}.

\begin{acknowledgements}
The computational resources and services used in this work were provided by the VSC (Flemish Supercomputer Center), funded by the Research Foundation – Flanders (FWO) and the Flemish Government. The authors acknowledge financial support from the Research Foundation Flanders (T.B., FWO grant G098320N). 
\end{acknowledgements}

\begin{das}
	The processed data that supports the findings of this study will be made available upon acceptance of the manuscript.
\end{das}

\section*{Appendix 1: Resolution Sensitivity}
\label{app:resolution}

To assess the sensitivity of our LES results to the grid resolution, we ran the three simulations with the strongest stability on a finer grid. These cases will be indicated as S6f, S7f and T4f. The grid size resolution is doubled with repsect to the cases in Table  \ref{tab:overview}, i.e. $\mathrm{\Delta}_x=\mathrm{\Delta}_y=1.56$ m in the horizontal directions and $\mathrm{\Delta}_z = 0.78$ m vertically. For case T4f, the domain height was halved to $L_z=200$ m in order to reduce computational cost. Given that both the momentum and heat flux profiles are constant above $h_\theta \approx 70$ m (see Fig. \ref{fig:profiles}), it is safe to assume that this does not affect the boundary layer itself. By contrast, the turbulence in cases S6 and S7 penetrates deeper into the free atmosphere (see Figs. \ref{fig:profiles} and \ref{fig:scaled_profiles}), prohibiting to reduce the domain height. 
We note that these simulations were initialized from instantaneous flow fields of the cases with coarser grid, interpolated towards the finer mesh. Due to the pseudo-spectral discretization, horizontal interpolation was achieved through zero-padding the Fourier transform as explained in e.g. \cite{McWilliams2023}. The fine-grid simulations were run for approximately 12 hours to allow the resolved small-scale turbulence to develop and reach a new thermal equilibrium state, after which the next 12.5 hours were used to collect statistics. Hence, the total runtime for each simulation was 24 physical hours.

\begin{figure*}[h]
	\centering
	\includegraphics[width=\textwidth]{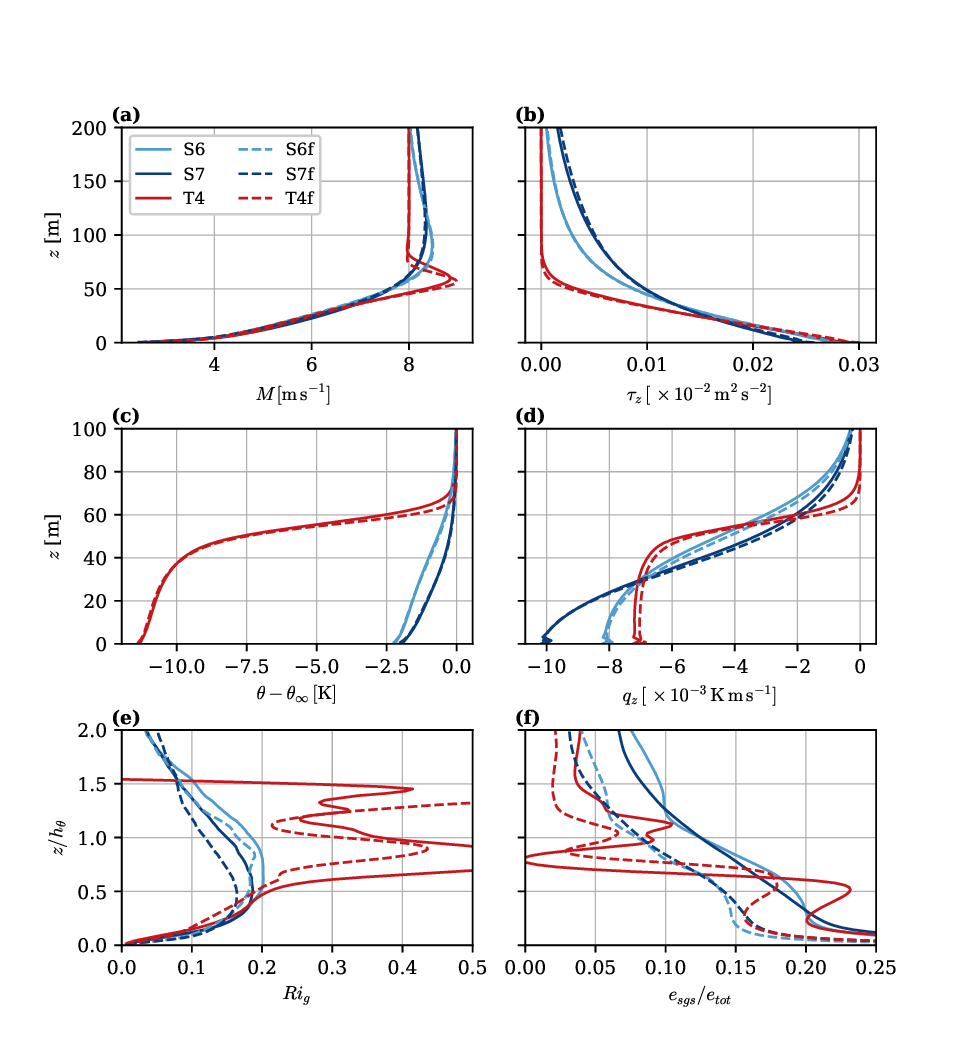}
	\caption{Comparison of simulations S6, S7 and T4 (full lines) from Table \ref{tab:overview} to simulations with doubled grid resolution (dashed lines). Subplots show mean wind speed (a), vertical momentum flux (b), potential temperature deviation (c), vertical heat flux (d), gradient Richardson number (e) and the ratio of subgrid to total TKE (f)}\label{fig:resolution_sensitivity}
\end{figure*}

Figure \ref{fig:resolution_sensitivity} shows the effect of increasing the grid resolution on the mean profiles of horizontal velocity (a), vertical momentum flux (b), potential temperature (c) and vertical heat flux (d). The differences between the fine and coarse grid cases appear rather limited, indicating that the grid-size used in this study is sufficient. The discrepancies are found to be largest for case T4f, which is also confirmed by comparing the global flow quantities in Table \ref{tab:resolution}. We find that for cases S6 and S7 the impact of doubling the grid resolution on $u_*, q_0$ and $h_\theta$ is less than 1.5\%, while in case T4f the surface heat flux and thermal boundary-layer height are reduced by approximately 2.5 and 9\%, respectively. The reduction of boundary-layer height with increasing grid resolution is a known issue in LES, as discussed elaborately by \cite{Maronga2022}. This is exemplified by the resolution sensitivity analysis by L19, which shows a 9\% increase of SBL depth in their VSBL case when the grid spacing is raised by a factor 1.5. 
Following \cite{Sullivan2016a}, we also compare the gradient Richardson number $Ri_g$ for the different grid resolutions in Fig. \ref{fig:resolution_sensitivity}e. This height-dependent stability characteristic is defined as $Ri_g(z) = N^2/S^2$, where $N$ is the Brunt-Vaisala frequency and $S = [(\partial u/\partial z)^2 + (\partial v/\partial z)^2]^{1/2}$ the mean shear. Both $N^2$ and $S^2$, and therefore $Ri_g$, are sensitive to mesh resolution. According to \cite{Sullivan2016a}, LES with (too) coarse resolution is unable to resolve small-scale vertical motions in the entrainment zone above the LLJ, because the Ozmidov Length becomes smaller than the grid spacing. As a result, $S^2$ decreases and $Ri_g$ sharply increases above the critical value of $~0.25$. We indeed observe a strong peak in $Ri_g$ around the top of the capping inversion ($z=h_\theta$) in case T4, while its magnitude is reduced in case T4f, related to reduced shear in the coarser-resolution case (not shown). For the two cases with high subsidence rate, S6 and S7, the critical Richardson number is not reached, allowing turbulence to penetrate higher into the atmosphere as mentioned in Sect. \ref{sec:results}. 

\cite{Matheou2014} proposed a criterion for grid convergence based on the resolved TKE, stating that 90\% of the TKE should be resolved for convergence of first-order statistics whereas 95\% is required for second-order statistics. Fig. \ref{fig:resolution_sensitivity}f displays the ratio between the SGS-TKE and the total TKE ($e_{tot} = (\tilde{u}_i\tilde{u}_i)/2 + e_{sgs}$). It is evident that the subgrid contribution reduces when a finer mesh is used. If we ignore the surface layer, the subgrid contribution is smaller than 20\% in cases S6 and S7 and smaller than 15\% in cases S6f and S7f. Hence, the requirement from \cite{Matheou2014} is not met, yet given the small differences in Figs. a-d and Table \ref{tab:resolution}, we still judge the grid resolution to be sufficient for the purpose of this study. Figure \ref{fig:resolution_sensitivity}f further shows that in case T4, the SGS-TKE (and thus the subrid flux) goes to zero due to the strong stability in the inversion layer, while in simulation T4f the subrid TKE does not vanish. This is another indication that the inversion layer in case T4 is under-resolved. However, we remark that replacing the data from this coarse-resolution case by that from T4f does not alter any of the qualitative findings reported in the body of this paper. In fact, the normalized temperature and heat flux profiles from case T4f collapse better on the other profiles from set T in Figs. \ref{fig:scaled_profiles}c and d (not shown).

\begin{table*}[h]
	\caption{Global flow properties of the high-resolution simulations S6f, S7f and T4f. The relative error is defined as $\delta[\chi] = (\chi_f-\chi_c)/\chi_c$, where $\chi_f$ refers to the value from the simulations with the finer grid and $\chi_c$ to the coarser-grid simulation results given in Table \ref{tab:overview}.}
	\label{tab:resolution}
	\begin{center}
		\begin{tabular}{lllllll}
			\hline
			Case      & $u_*$ (m s$^{-1}$) & $\delta[u_*]\, (\%)$ & $q_0$ (K m s$^{-1}$) &  $\delta[q_0]\, (\%)$  & $h_\theta$ (m) &  $
			\delta[h_\theta] \,(\%)$ \\
			\hline
			S6f & 0.169 & 0.41 & $-8.15 \times 10^{-3}$ & -0.98 & 95.7 & -0.41\\
			S7f & 0.161 & 1.38 & $-10.2 \times 10^{-3}$ & 0.00  & 80.9 & -1.43\\
			T4f & 0.174 & 1.13 & $-7.05 \times 10^{-3}$ & -2.47 & 67.6 & -8.95\\
		\end{tabular}
		
	\end{center}
\end{table*}

\section*{Appendix 2: Comparison to the Weakly Stable Boundary Layer Case from \cite{VanderLinden2019}}
\label{app:vanderlinden}
Here, we compare the WSBL simulation from L19 to the results obtained with the present SP-Wind code. The full case setup is provided in tables 1 and 2 of L19, yet we shortly point out some key parameters and relevant differences between the WSBL-B24 and WSBL-L19 simulations. They mention they use a subsidence velocity of 4 mm s$^{-1}$ at $z=100$ m that was linearly interpolated to zero at the surface, corresponding to $\mathrm{\Gamma}_w = \mathrm{d}w_s/\mathrm{d}z = 4~\times~10^{-5}$~s$^{-1}$. The geostrophic wind speed ($G = 12$ m s$^{-1}$), global temperature difference ($\mathrm{\Delta}\theta = 25$ K), and reference temperature ($\theta_r = 235$ K) deviate from the the other simulations in the present study. The roughness length for momentum is $z_{0m}=10^{-3}$ m, while the roughness length for temperature is one order of magnitude smaller. The domain size is $336 \times 336 \times 168$ m. 

The main difference between the MicroHH code used by L19 and the SP-Wind code used here, is the discretization scheme. While they apply second-order finite differences for advection, SP-Wind is a pseudospectral code with Fourier-decomposition in the horizontal directions and a fourth-order finite difference scheme in the vertical direction. Motivated by this difference in discretization, case WSBL-B24 was computed with a grid spacing of $\mathrm{\Delta}_x=\mathrm{\Delta}_y = 1.31$ m and $\mathrm{\Delta}_z=0.7$ m, larger than the isotropic grid with $\mathrm{\Delta}=0.7$ m used by L19. Furthermore, L19 employ a Smagorinsky-Lily type eddy-viscosity model with a wall correction for the mixing length to calculate the subgrid fluxes, whereas the present simulations utilized a closure based on SGS-TKE (see Sect. \ref{subsec:numericalmethods}).

The initialization procedure used in this study is outlined in Sect. \ref{subsubsec:initialization}. In case WSBL-B24 we used an initial temperature profile that increased linearly from $\theta_s$ to $\theta_s + 25$ K at $h_i = 50$ m, whereas L19 initialized their simulations with a constant temperature and cooled for 25 hours until the final surface temperature was reached. Subsequently, they continued the simulation for 5 hours and collected data from the final hour. In contrast, case WSBL-B24 was run for 32 physical hours in total, and the averaging window was a full inertial period ($\approx 12.5$ h) in which the heat flux due to subsidence was equal to the surface heat flux. 

\begin{figure}[h]
	\centering
	\includegraphics{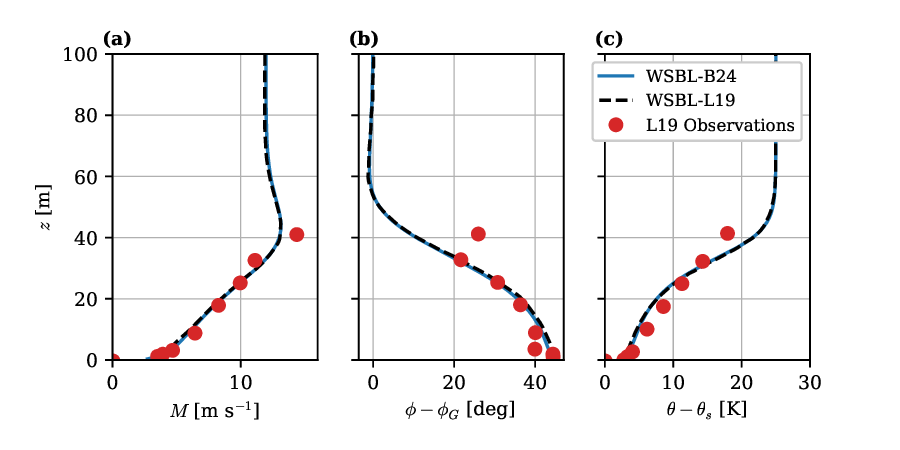}
	\caption{Comparison of mean profiles of wind speed (a),relative wind direction (b) and potential temperature (c) for WSBL cases from the present study (blue lines) and L19 (black dashed lines). Data from from Fig. 5 in L19 was extracted using WebPlotDigitizer (apps.automeris.io/wpd/). Red circles indicate observed values from a measurement campaign in Antarctica as reported in L19}\label{fig:L19_profiles}
\end{figure}

Figure \ref{fig:L19_profiles} compares the mean profiles of horizontal velocity, flow angle and temperature for cases WSBL-L19 and WSBL-B24. An online data extractor tool was used to obtain the data from Fig. 5 in L19. These first-order profiles match rather well. For reference, the observational results from the Dome C site in Antarctica, where L19 compare their data to, is also included. The LES results agree well with these field measurements, except in the highest grid point as discussed by L19. The surface fluxes of the WSBL-L19 and WSBL-B24 are reported in Table \ref{tab:overview}, where the largest difference is a 4\% deviation of the surface heat flux $q_0$. We further point out that the boundary-layer height calculated from the momentum flux profiles according to the first method in Sect. \ref{subsec:sbl_depth} is $h_\tau=47.16$ m, equal to $h = 47.2$ m reported by L19. Overall, we conclude that the correspondence among the simulation results is good, especially considering the differences in numerical methodology outlined above.

\section*{Appendix 3: A Short Note on Validity of Monin--Obukhov Similarity Theory as a Boundary Condition in Large-Eddy Simulations with subsidence}
\label{app:most_validity}
As there is a clear dependency of the dimensionless gradients $\phi_{m/h}$ on subsidence rate (see Fig. \ref{fig:MOST_surface}), the question arises whether the use of MOST for parametrization of surface fluxes is justified in the presence of subsidence. The simplest answer would be that the subsidence velocity approaches zero close to the surface, and is therefore negligible in the surface layer \citep[e.g.][]{Carlson1985}.
We attempt to support this statement through a scaling argument, to illustrate that the heat flux due to subsidence is indeed negligible compared to the surface heat flux in the present simulation setup. Assuming MOST is valid and the dimensionless temperature gradient is linear in $z/L_O$, the (dimensional) temperature gradient is given by:
\begin{equation}
	\frac{\partial \theta}{\partial z} = \frac{\theta_*}{\kappa z} \left(\alpha_h + \beta_h \frac{z}{L_O} \right). 
\end{equation}
Then the subsidence heating in the surface layer can be written as:
\begin{equation}
	Q_{\mathrm{subs}} = \mathrm{\Gamma}_w z \frac{\partial \theta}{\partial z} = \frac{\mathrm{\Gamma}_w\theta_*}{\kappa} \left(\alpha_h + \beta_h \frac{z}{L_O} \right).
	\label{eq:Qsubs}
\end{equation}
Since MOST is only applied in the first layer of grid cells above the surface, we integrate \eqref{eq:Qsubs} from $z_0$ up to the first grid level $z_1$ to find the effective subsidence heat flux in the first grid cell:
\begin{equation}
	q_{\mathrm{subs}}=\int_{z_0}^{z_1}Q_{\mathrm{subs}} \mathrm{d}z = \frac{\mathrm{\Gamma}_w\theta_*}{\kappa} \left(\alpha_h + \frac{\beta_h}{2}\frac{\mathrm{\Delta} z_1}{L_O}\right)\mathrm{\Delta} z_1,
	\label{eq:qsubs}
\end{equation}
where we assumed that $z_1 \gg z_0$ such that $z_1-z_0$ can be replaced by the size of the first grid cell $\mathrm{\Delta} z_1$. Furthermore, it should be safe to assumed that $\mathrm{\Delta} z_1 \ll L_O$ and $\beta_h/2 \sim O(1)$, allowing the second term in Eq. \eqref{eq:qsubs} to be dropped. Then the ratio between the subsidence heat flux and the surface heat flux ($q_0 = -u_*\theta_*$) can be estimated as:
\begin{equation}
	R_s = \frac{\left| q_{\mathrm{subs}}\right|}{\left|q_0\right|} = \frac{\mathrm{\Gamma}_w \mathrm{\Delta} z_1}{\kappa u_*}.
	\label{eq:R_q}
\end{equation}
To estimate this ratio in the simulations of the present study, we take $\mathrm{\Gamma}_w \approx 10^{-4}$ s$^{-1}$ , $\mathrm{\Delta} z_1 \approx 1$ m and $u_* \approx 0.1$~m~s$^{-1}$, resulting in $R_s \approx 2.5 \times 10^{-3} \ll 1$. With this, we prove that the heat flux due to subsidence in the first grid level is indeed negligible in comparison to the surface heat flux. Note that Eq. \eqref{eq:R_q} in combination with the estimation of $u_*$ given by Eq. \eqref{eq:GDL} can be used to determine the grid spacing that is required to satisfy the condition that $R_q \ll 1$ a priori.

\bibliographystyle{spbasic_updated}     
\bibliography{library.bib}

\end{document}